\documentclass[aps, twocolumn, nobalancelastpage, amsmath, amssymb,
superscriptaddress, nofootinbib]{revtex4-1}

\usepackage{graphics}      
\usepackage{graphicx}      
\usepackage{longtable}     
\usepackage{bm}            
\usepackage{natbib}
\usepackage{physics}
\usepackage{xcolor}
\usepackage[caption=false]{subfig}
\usepackage{dcolumn}
\usepackage{hyperref}

\newcommand{\Center}[1]{\multicolumn{1}{c}{#1}}
\newcommand{\Ion}[3]{${}^{#1}\mathrm{#2}^{#3}$}
\newcommand{\Neg}[1]{-#1}
\newcommand{\Pos}[1]{\phantom{-}#1}
\newcommand{\Change}[1]{\textcolor{black}{#1}}

\newcommand{\Za}{{Z\alpha}}
\newcommand{\balpha}{\bm{\alpha}}

\newcommand{\Dmatrix}[4]{
  \left(
    \begin{array}{cc}
      #1  & #2   \\
      #3  & #4   \\
    \end{array}
  \right)
}

\begin{document}

\title{QED corrections to the $g$~factor of Li- and B-like ions}

\author{H. Cakir}
\email{halil.cakir@mpi-hd.mpg.de}
\affiliation{Max-Planck-Institut f\"{u}r Kernphysik, Heidelberg, Germany}
\author{V. A. Yerokhin}
\affiliation{Max-Planck-Institut f\"{u}r Kernphysik, Heidelberg, Germany}
\affiliation{Peter the Great St.~Petersburg Polytechnic University,
  St.~Petersburg, Russia}
\author{N. S. Oreshkina}
\affiliation{Max-Planck-Institut f\"{u}r Kernphysik, Heidelberg, Germany}
\author{B. Sikora}
\affiliation{Max-Planck-Institut f\"{u}r Kernphysik, Heidelberg, Germany}
\author{I. I. Tupitsyn}
\affiliation{St. Petersburg State University, St.~Petersburg, Russia}
\author{C. H. Keitel}
\affiliation{Max-Planck-Institut f\"{u}r Kernphysik, Heidelberg, Germany}
\author{Z. Harman}
\affiliation{Max-Planck-Institut f\"{u}r Kernphysik, Heidelberg, Germany}

\date{March~2,~2020}

\begin{abstract}
  QED corrections to the $g$~factor of Li-like and B-like ions in a wide range
  of nuclear charges are presented.  Many-electron contributions as well as
  radiative effects on the one-loop level are calculated.  Contributions
  resulting from the interelectronic interaction, the self-energy effect, and
  most of the terms of the vacuum-polarization effect are evaluated to all
  orders in the nuclear coupling strength~$Z\alpha$.  Uncertainties resulting
  from nuclear size effects, numerical computations, and uncalculated effects
  are discussed.
\end{abstract}

\maketitle

\section{Introduction}

Precision studies of $g$~factors of highly charged ions (HCI) provide a unique
possibility for testing fundamental theories.  Penning-trap experiments
employing the continuous Stern-Gerlach effect achieve nowadays a high
precision, and are advancing towards heavy ions, in which effects of quantum
electrodynamics (QED) are most relevant. The $g$~factor of hydrogen-like
silicon ($Z=14$) has been determined with a $5\cdot 10^{-10}$ fractional
uncertainty~\cite{Sturm2011, Sturm2013}, allowing to scrutinize bound-state QED
theory (see e.g.~\cite{Pachucki2004, Pachucki2005, Karshenboim2002, Lee2005,
  Yerokhin2002, Yerokhin2004, Shabaev2002-2, Beier2000-1, Beier2000}).
Recently, the evaluation of two-loop terms of order $(Z\alpha)^{5}$ (with $Z$
being the atomic number and $\alpha$ the fine-structure constant) has been
finalized~\cite{Czarnecki2018} (see also \cite{Pachucki2017}), increasing the
theoretical accuracy especially in the low-$Z$ regime.  First milestones have
been reached in the calculation of
two-loop corrections for non-perturbative Coulomb fields, i.e. for larger
values of $Z\alpha$~\cite{Yerokhin2013-1, Sikora2020}.

The high accuracy which can be achieved on the experimental as well as
theoretical side also enables the determination of fundamental physical
constants such as the electron mass~\cite{Sturm2014, Koehler2015,
  Zatorski2017}.  However, QED tests as well as the extraction of fundamental
constants may be limited by nuclear effects~\cite{Glazov2002, Beier2000,
  Nefiodov2002, Zatorski2012}.  Since the nuclear parameters entering the
nuclear corrections are not always sufficiently well known, these corrections
may be associated with large uncertainties, and thus set a natural limit to the
accuracy of the theoretical $g$~factor.

The extension of experiments to the heaviest ions, including Pb${}^{81+}$ and
U${}^{91+}$, is expected in the forthcoming years by the use of the ALPHATRAP
Penning-trap setup~\cite{Sturm2013-1} and the HITRAP facility~\cite{Quint2001,
  Kluge2008}.  Measurements with these systems are anticipated to provide an
alternative determination of the value of $\alpha$ \cite{Shabaev2006,
  Yerokhin2016, Yerokhin2016-1}.  In Ref.~\cite{Shabaev2006}, a specific (or
weighted) difference of the $g$~factors of heavy H- and B-like ions with the
same nuclear species was put forward.  It was demonstrated that the theoretical
uncertainty of the nuclear finite size effects in this difference can be
suppressed down to $4\times10^{-10}$ for very heavy ions such as Pb, which was
several times smaller than the theoretical uncertainty of the $g$~factor due to
$\alpha$ at that time.  In Ref.~\cite{Yerokhin2016, Yerokhin2016-1} a specific
difference of the $g$-factors of low-$Z$ H- and Li-like ions was proposed, for
which an even stronger suppression of nuclear effects and their uncertainties
can be achieved, leading to an accuracy competitive with the current value of
$\alpha$.  A calculation of the nuclear polarization effect extended to Li- and
B-like ions showed that these terms can also be largely suppressed in a
specific difference of the $g$~factors for two different charge states of the
same element~\cite{Volotka2014-1}.

Motivated by these prospects, in the current paper, we calculate the
ground-state $g$~factor of Li- and B-like HCI.  Our results for Li-like ions
confirm previous calculations.  \Change{We extend the computations for B-like
Ar${}^{13+}$ presented in \cite{Arapoglou2019} for a range of elements across
the periodic table, and describe them in detail in the current
  manuscript}.  The one-electron self-energy term is calculated with an
improved numerical accuracy.  The vacuum polarization screening diagrams are
evaluated, and self-energy screening is estimated using effective screening
potentials.  Electron correlation effects are taken into account by exact QED
methods up to order~$1/Z$, and higher-order terms are extracted from
large-scale relativistic configuration interaction calculations.

This paper is organized as follows. In Section~\ref{sec:relativistic}, we
discuss relativistic and electron-correlation contributions to the
bound-electron $g$~factor.  In Sections~\ref{sec:self_energy} and
\ref{sec:vacuum_polarization}, we describe our computations of the self-energy
and vacuum-polarization contributions, respectively.  In
Section~\ref{sec:results}, we tabulate and discuss our computations of the
contributions to the bound-electron $g$~factor and provide concluding remarks.
We use relativistic units ($\hbar = c = m_{\mathrm{e}} = 1$) and the Heaviside
charge unit ($\alpha = e^{2} / (4\pi)$, $e < 0$).

\section{Relativistic $g$ factor}
\label{sec:relativistic}

The Zeeman shift linear in the magnetic field of an energy level of an atom
with a spinless nucleus is parameterized in terms of the $g$~factor of the atom
by the equation
\begin{equation}
  \label{eq:1}
  \Delta{E} = g \, \mu_{\mathrm{B}} \langle \bm{J} \cdot \bm{B} \rangle \,,
\end{equation}
where $\Delta{E}$ is the energy shift, $\bm{J}$ is the operator of the total
angular momentum, $\bm{B}$ is the external magnetic field,
$\mu_{\mathrm{B}} = |e|/2$ denotes the Bohr magneton, and $g$ is the
$g$~factor.  The $g$~factor is determined by computing the Zeeman energy
splitting and solving Eq.~(\ref{eq:1}) for $g$.

The relativistic interaction of an electron with the homogeneous external
magnetic field is given by
\begin{equation}
  \label{eq:2}
  V_{\mathrm{mag}}(\bm{r}) = -e \bm{\alpha} \cdot \bm{A}(\bm{r}) \,,
\end{equation}
where $\bm{A}$ is the vector potential
$\bm{A}(\bm{r}) = \left(\bm{B} \times \bm{r}\right) / 2$.  Assuming that the
magnetic field is directed along the $z$~axis, $V_{\mathrm{mag}}$ reduces to
\begin{equation}
  \label{eq:3}
  V_{\mathrm{mag}}(\bm{r}) = \frac{|e|B_{z}}{2} \, \left(\bm{r} \times
    \bm{\alpha}\right)_{z} \,.
\end{equation}

An \emph{ab initio} QED theory of the $g$~factor of an atom can be formulated
e.g. within the two-time Green's function formalism~\cite{Shabaev2002}.  Within
this formalism, the Zeeman energy splitting is calculated.

\subsection{Dirac value and nuclear size contribution}

\begin{figure}
  \centering
  \includegraphics[width=0.15\textwidth]{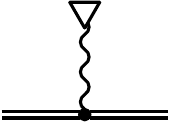}
  \caption{The Feynman diagram representing the leading contribution to the
    bound-electron $g$~factor.  Double lines represent electrons in the
    electric field of the nucleus and a wavy line with a triangle represents an
    interaction with the external magnetic field.}
  \label{fig:dirac}
\end{figure}

The leading contribution to the $g$~factor of an alkali-like atom arises
through the interaction of the valence electron with the external magnetic
field.  The corresponding Feynman diagram is depicted in Fig.~\ref{fig:dirac}.
Within the approximation of non-interacting electrons, contributions resulting
from the interaction of the closed-shell core electrons with the external
magnetic field cancel in the final sum, since electrons with opposite spin
projections induce contributions of the same magnitude but of opposite sign.
Therefore, only the contribution of the valence electron remains.  For this
reason, the $g$~factor of the whole alkali-like atom is often termed as the
bound-electron $g$~factor (assuming that of the valence electron).

The leading (Dirac) contribution to the bound-electron $g$~factor of an
alkali-like atom with the valence state characterized by the quantum numbers
$n$ and $\kappa$ is
\begin{equation}
  \label{eq:4}
  g_{\mathrm{D}} = \frac{2\kappa}{j(j+1)} \int_{0}^{\infty} dr \, r
  G_{n\kappa}(r) F_{n \kappa}(r) \,,
\end{equation}
where $j = |\kappa| - 1/2$ is the total angular momentum quantum number, and
the functions $G_{n \kappa}$ and $F_{n \kappa}$ are the radial components of
the electronic wave function
\begin{equation}
  \label{eq:5}
  \psi_{n \kappa m}(\bm{r})
  = \frac{1}{r}
  \begin{pmatrix} 
    G_{n\kappa}(r) \Omega_{\kappa m}(\bm{n}) \\
    i F_{n\kappa}(r) \Omega_{-\kappa m}(\bm{n})
  \end{pmatrix} \,,
\end{equation}
where $\Omega_{\kappa m}(\bm{n})$ are spherical spinors.

For a point-like nucleus, the integral in Eq.~(\ref{eq:4}) can be evaluated
analytically, with the result \cite{Breit1928, Zapryagaev1979}
\begin{equation}
  \label{eq:6}
  g_{\mathrm{D}}(\mathrm{pnt}) = \frac{\kappa}{2j(j+1)} \left(2 \kappa
    \varepsilon_{n \kappa} - 1\right) \,,
\end{equation}
where $\varepsilon_{n \kappa}$ is the Dirac energy of the reference state.  In
particular, for the $2s$ and $2p_{1/2}$ states relevant for this work, they are
\begin{equation}
  \label{eq:7}
  \varepsilon_{2s} = \varepsilon_{2p_{1/2}} = \sqrt{\frac{1 + \gamma}{2}}
\end{equation}
where $\gamma = \sqrt{1 - (Z\alpha)^2}$.

The nuclear size correction to the point-nucleus Dirac value is determined as
the difference of Eq.~(\ref{eq:4}) evaluated numerically for an extended
nuclear charge distribution and the point-nucleus result of Eq.~(\ref{eq:6}).
We use the homogeneously charged sphere as the model for an extended nucleus
with the RMS~radii taken from Ref.~\cite{Angeli2013}.  \Change{We estimate the
  dependence on the model by also using the two-parameter Fermi distribution
  and find it to be insignificant compared to the uncertainties associated with
  other contributions.}

\subsection{First-order interelectronic interaction}

Interactions among the electrons in a multi-electron ion result in a
contribution to the bound-electron $g$~factor.  These interactions can be
classified according to the number of exchanged photons and the associated
perturbation parameter is~$1/Z$.

\begin{figure}
  \centering
  \includegraphics[width=0.23\textwidth]{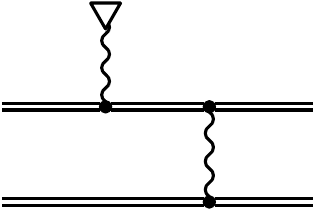}
  \caption{A typical Feynman diagram representing the one-photon
    interelectronic-interaction contribution to the bound-electron $g$~factor.
    Only diagrams where one of the electrons is the valence electron and the
    other one a core electron contribute.}
  \label{fig:inter}
\end{figure}

We compute the leading one-photon exchange contribution, which corresponds to
the first-order perturbation correction in the parameter~$1/Z$.  A typical
contributing diagram is depicted in Fig.~\ref{fig:inter}.  As seen in the
figure, the computation of the one-photon exchange contribution reduces the
many-electron problem to a two-electron one, where one of the core electrons
interacts with the valence electron in addition to the interaction with the
external magnetic field.  Analogous contributions from the exchange between
core electrons vanish, again, identically after summing over the momentum
projections of the closed-shell core states.

The one-photon exchange contribution to the Zeeman shift of an energy level can
be expressed as the sum of an \emph{irreducible} and a \emph{reducible} part;
the corresponding formulas were derived in Ref.~\cite{Shabaev2002-1}.  The
irreducible part arises from the first-order perturbative correction to the
bound-electron wave function,
\begin{equation}
  \label{eq:8}
  \ket{\delta{a}} = \sum_{n}^{\varepsilon_{n} \neq \varepsilon_{a}}
  \frac{\bra{n}V_{\mathrm{mag}}\ket{a}}{\varepsilon_{a} - \varepsilon_{n}}
  \ket{n} \,,
\end{equation}
where $a$ is either a core-electron state~$c$ or a valence-electron state~$v$
and the summation label~$n$ runs over the whole electronic spectrum including
all bound states except the reference state~$a$.  The irreducible contribution
is then given by
\begin{multline}
  \label{eq:9}
  \Delta{E}_{\mathrm{int, irr}}^{(1)} = 2 \sum_{c}
  \Big(\bra{vc}I(0)\ket{\delta{v}c} - \bra{cv}I(\Delta_{vc})\ket{\delta{v}c} \\
  + \bra{vc}I(0)\ket{v\delta{c}} - \bra{cv}I(\Delta_{vc})\ket{v\delta{c}}\Big)
  \,,
\end{multline}
where the summation is carried out over all core states,
$\Delta_{vc} = \varepsilon_{v} - \varepsilon_{c}$ is the difference between the
Dirac energy levels of the valence and core electrons, $I$ is the operator of
the electron-electron interaction,
\begin{equation}
  \label{eq:10}
  I(\omega, \bm{r}_{1}, \bm{r}_{2}) = e^{2} \alpha_{1}^{\mu} \alpha_{2}^{\nu}
  D_{\mu\nu}(\omega, \bm{r}_{12}) \,,
\end{equation}
${\alpha}^{\mu} = (1, \balpha)$ are the Dirac matrices, $D_{\mu\nu}$ is the
photon propagator, and $\bm{r}_{12} = \bm{r}_{1} - \bm{r}_{2}$.

The reducible contribution arises from first-order perturbations of the
energies of the core and valence electrons by the magnetic interaction,
$\varepsilon_{a} \mapsto \varepsilon_{a} + \bra{a}V_{\mathrm{mag}}\ket{a}$.  It
is given by
\begin{multline}
  \label{eq:11}
  \Delta{E}_{\mathrm{int, red}}^{(1)} = \sum_{c}
  \bra{cv}I'(\Delta_{vc})\ket{vc} \times \\
  \times \Big(\bra{c}V_{\mathrm{mag}}\ket{c} -
  \bra{v}V_{\mathrm{mag}}\ket{v}\Big) \,,
\end{multline}
where the prime on $I'(\omega)$ denotes the derivative with respect to
$\omega$.

For the numerical computation of the one-photon exchange correction we solve
the radial Dirac equation using basis sets constructed from $B$~splines within
the dual kinetic balance~(DKB) approach~\cite{Johnson1988, Shabaev2004}.  This
approach is particularly suited for the computation of spectral sums as in
Eq.~(\ref{eq:8}).  In our numerical treatment of the radial Dirac equation we
take the nuclear size into account by using a homogeneously charged sphere as a
nucleus with RMS~radii taken from Ref.~\cite{Angeli2013}.  \Change{The
  contributions are calculated using the Feynman and Coulomb gauges in order to
  estimate the numerical uncertainty.} We present our result in
Table~\ref{tab:inter}.  Our calculations of the one-photon exchange correction
reproduce previous results obtained in Ref.~\cite{Shabaev2002-1} for Li-like
ions and Ref.~\cite{Glazov2013, Agababaev2018} for B-like ions.

\begin{table}
  \caption{First-order interelectronic-interaction contribution to the
    bound-electron $g$~factor of the ground state of Li- and B-like
    ions. \Change{The uncertainties account for uncertainties in the nuclear
      RMS radii and numerical errors}.}
  \begin{ruledtabular}
    \begin{tabular}{lll}
                   & \multicolumn{2}{c}{Electron correlation, $(1/Z)^{1}$}   \\
      \Center{$Z$} & \Center{Li-like}           & \Center{B-like}            \\
      \hline                                                                 \\[-5pt]
      18           &  0.000\,414\,450\,489\,(3) &  0.000\,657\,531\,117\,(1) \\
      20           &  0.000\,461\,147\,896\,(3) &  0.000\,731\,996\,913\,(1) \\
      24           &  0.000\,555\,185\,23\,(1)  &  0.000\,882\,350\,695\,(5) \\
      32           &  0.000\,746\,458\,66\,(1)  &  0.001\,190\,274\,990\,(5) \\
      54           &  0.001\,306\,216\,8\,(4)   &  0.002\,118\,178\,3\,(3)   \\
      82           &  0.002\,148\,290\,(1)      &  0.003\,654\,888\,(2)      \\
      92           &  0.002\,509\,828\,(7)      &  0.004\,393\,71\,(1)       \\
    \end{tabular}
  \end{ruledtabular}
  \label{tab:inter}
\end{table}

\subsection{Higher-order interelectronic interaction}

Electron correlation effects beyond the first-order approximation in~$1/Z$,
described in the previous subsection, were taken into account by means of a
relativistic configuration interaction Dirac-Fock-Sturm~(CI-DFS) approach,
employing Dirac-Hartree-Fock orbitals for the occupied states and relativistic
Sturmian orbitals for the virtual electronic states as in
Ref.~\cite{Shchepetnov2015}.  The contribution of the negative-energy part of
the Dirac spectrum, which was found to be relevant in case of $g$~factors of
Li-like ions in Ref.~\cite{Wagner2013}, is also significant in the case of
B-like ions, and the corresponding states were also described with Sturmian
orbitals.

The one-photon exchange correction results in the Breit approximation were used
to monitor the convergence of the CI calculations when systematically extending
the Sturmian basis set in the latter.  The configurations obtained by single,
double, and triple excitations of the ground state were included in the
calculation.  The one-electron functions were chosen up $n=10$ and $l=5$,
leading to the total number of over $100000$ configurations.  The theoretical
uncertainty was estimated as twice the difference of the results using
the largest and the second largest Sturmian basis sets.

\section{Self energy}
\label{sec:self_energy}

\subsection{One-electron self energy}

To the zeroth order in~$1/Z$, we can ignore the presence of the core electrons
and evaluate the self-energy~(SE) correction assuming the reference state being
the hydrogenic Dirac state of the valence electron~$v$.

\begin{figure}
  \centering
  \subfloat[\label{fig:se_a}]{\includegraphics[width=0.15\textwidth]{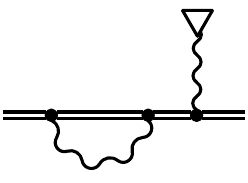}}
  \subfloat[\label{fig:se_b}]{\includegraphics[width=0.15\textwidth]{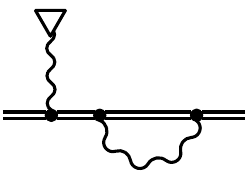}}
  \subfloat[\label{fig:se_c}]{\includegraphics[width=0.15\textwidth]{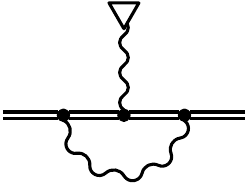}}
  \caption{Feynman diagrams representing the self-energy contribution to the
    bound-electron $g$~factor.}
  \label{fig:se}
\end{figure}

The SE~contribution to the energy shift of the hydrogenic state~$v$ in the
presence of a perturbing potential $V_{\mathrm{mag}}$ is graphically
represented by the Feynman diagrams shown in Fig.~\ref{fig:se}.  The general
expression for the SE~correction can be conveniently split into three
parts~\cite{Shabaev2002},
\begin{equation}
  \label{eq:12}
  \Delta{E}_{\mathrm{SE}} = \Delta{E}_{\mathrm{SE, irr}} +
  \Delta{E}_{\mathrm{SE, red}} + \Delta{E}_{\mathrm{SE, ver}} \,,
\end{equation}
which are referred to as the \emph{irreducible}, the \emph{reducible}, and the
\emph{vertex} contribution, respectively.

The irreducible contribution is induced by a part of the diagrams in
Fig.~\ref{fig:se_a} and Fig.~\ref{fig:se_b} that can be expressed in terms of
the first-order perturbation of the reference-state wave function by
$V_{\mathrm{mag}}$ given in Eq.~(\ref{eq:8}),
\begin{equation}
  \label{eq:13}
  \Delta{E}_{\mathrm{SE, irr}} = 2 \bra{v}\gamma^{0}
  \widetilde{\Sigma}(\varepsilon_{v})\ket{\delta{v}} \,,
\end{equation}
where $\widetilde{\Sigma} = \Sigma - \delta{m}$, $\delta{m}$ is the one-loop
mass counterterm, and $\Sigma$ is the one-loop SE~operator,
\begin{multline}
  \label{eq:14}
  \Sigma(\varepsilon, \bm{r}_{1}, \bm{r}_{2}) = 2 i \alpha \gamma^{0}
  \int_{C_{F}} d\omega \, \alpha_{\mu} G(\varepsilon - \omega, \bm{r}_{1},
  \bm{r}_{2}) \times \\
  \times \alpha_{\nu} D^{\mu\nu}(\omega, \bm{r}_{12}) \,.
\end{multline}
Here, $G$ denotes the Dirac Coulomb Green function
$G(\varepsilon) = [\varepsilon - \mathcal{H} (1 - i 0^{+})]^{-1}$,
$\mathcal{H}$ is the Dirac Coulomb Hamiltonian, and $C_{F}$ denotes the
standard Feynman integration contour.

The reducible contribution is induced by a part of the diagrams in
Fig.~\ref{fig:se_a} and Fig.~\ref{fig:se_b} that can be expressed in terms of
the first-order perturbation of the reference-state energy.  It reads
\begin{equation}
  \label{eq:15}
  \Delta E_{\mathrm{SE, red}} = \bra{v}\gamma^{0} \Sigma'(\varepsilon_{v})\ket{v}
  \bra{v}V_{\mathrm{mag}}\ket{v} \,,
\end{equation}
where the prime on $\Sigma'(\varepsilon)$ denotes the derivative with respect
to $\varepsilon$.

Finally, the vertex contribution is induced by the diagram in
Fig.~\ref{fig:se_c}.  It can be expressed as
\begin{multline}
  \label{eq:16}
  \Delta E_{\mathrm{SE, ver}} = \frac{i}{2\pi} \times \\
  \times \int_{-\infty}^{\infty} d\omega \, \sum_{n_1, n_2}
  \frac{\bra{n_{1}}V_{\mathrm{mag}}\ket{n_{2}}
    \bra{vn_{2}}I(\omega)\ket{n_{1}v}}{(\varepsilon_{v} - \omega - u
    \varepsilon_{n_1}) (\varepsilon_{v} - \omega - u \varepsilon_{n_2})} \,,
\end{multline}
where $u = 1 - i 0^{+}$ and the summations over $n_{1}$ and $n_{2}$ involve
both the positive-energy discrete and continuous spectra and the
negative-energy continuous spectrum.

Calculations of the SE~correction to the $g$~factor for the hydrogenic states
are rather complicated but well established by now.  For the point-nucleus
case, the most accurate computations were performed in
Refs.~\cite{Yerokhin2004, Yerokhin2010, Yerokhin2017}.  The finite nuclear size
correction was computed in Ref.~\cite{Yerokhin2013}.  In the present work, we
employ the numerical approach developed in those studies and extend the
previous calculations to the case of the $2p_{1/2}$ reference state (required
for B-like ions) and nuclear charges $Z > 12$, which has not been reported in
the literature.

\subsection{Screened self-energy}

The interaction of the valence electron with the core electrons modifies the
SE~effect and the resulting energy shift is known as the \emph{screened}
SE~correction.  It is suppressed by a small parameter~$1/Z$ as compared to the
leading SE~contribution.

A rigorous QED calculation of the screened SE~correction to the $g$~factor has
been performed in Refs.~\cite{Volotka2009, Volotka2014} for four Li-like ions
(with $Z = 14$, $20$, $82$, and~$92$).  This was a very difficult calculation,
which has not been so far extended to any other ion.  Two less sophisticated
methods exist in the literature for an approximate treatment of the screening
of the SE~corrections.  One method, used in Ref.~\cite{Glazov2006}, evaluates
the one-electron SE~correction in the presence of an additional screening
potential resulting from the interaction with the core electrons.  Another
method~\cite{Glazov2004} describes the electron self-energy by the anomalous
magnetic moment, which yields results complete to order~$(\Za)^{2}$ for
$s$~states.  In the following, we address these two methods in turn.

\subsubsection{Screening-potential approximation}
\label{sec:scr}

Within the screening-potential approximation, we consider the electron in the
combined field of the nucleus and an additional screening potential
$V_{\mathrm{scr}}$ that partly accounts for the interaction of the valence
electron with the core electrons.  The simplest choice of $V_{\mathrm{scr}}$ is
the core-Hartree~(CH) potential defined as
\begin{equation}
  \label{eq:17}
  V_{\mathrm{CH}}(r) = 4 \pi \alpha \int_{0}^{\infty} dr' \, r'^{2}
  \frac{\varrho_{\mathrm{core}}(r')}{r_{>}} \,,
\end{equation}
where $r_{>}$ is the larger one of $r$ and $r'$, and $\varrho_{\mathrm{core}}$
denotes the combined radial charge density of the core electrons in units of
the elementary charge.  In the present work we use also two other choices of
the screening potential, namely, the Kohn-Sham~(KS) potential and the local
Dirac-Fock~(LDF) potential, which are described in detail in
Ref.~\cite{Yerokhin2007}.  All screening potentials are constructed with the
Dirac-Fock wave functions.

Within the screening-potential approximation, the screened SE~effect is
obtained by evaluating the SE~correction according to
Eqs.~(\ref{eq:12})-(\ref{eq:16}) for the valence electron in the combined field
of the nucleus and $V_{\mathrm{scr}}$ and subtracting the SE~correction in the
nuclear field.  In this calculation, we generalized the numerical approach of
Ref.~\cite{Yerokhin2004} for computing the SE~correction to the $g$~factor to
the case of an arbitrary binding potential.  We used the Green's-function
technique, with the Green's function of the Dirac equation in a general
(asymptotically Coulomb) potential being computed by the method described in
the Appendix of Ref.~\cite{Yerokhin2011}.

\subsubsection{Anomalous magnetic moment approximation}

The second method for the approximate treatment of the SE~correction is based
on the nonrelativistic expansion.  As demonstrated in Ref.~\cite{Hegstrom1973},
to leading order in~$\Za$, which is here~$\propto\!(\Za)^{2}$, the
SE~correction to the bound-electron $g$~factor of an $s$~state is induced by
the interaction of the anomalous magnetic moment~(amm) of the electron with the
electric and magnetic field in the atom.  The interaction can be represented by
the following effective Hamiltonian,
\begin{equation}
  \label{eq:18}
  H_{\mathrm{amm}} = \sum_{j} \big(H_{1}(j) + H_{2}(j)\big) + \sum_{j \ne k}
  H_{3}(j, k) \,,
\end{equation}
where $j$ and $k$ numerate electrons and
\begin{align}
  \label{eq:19}
  H_{1}(j) &\ = a_{e} \mu_{B} \beta_{j} \bm{B} \cdot \bm{\Sigma}_{j} \,, \\
  \label{eq:20}
  H_{2}(j) &\ = a_{e} \frac{\Za}{2} (-i) \, \beta_{j} \frac{\bm{\alpha}_{j}
             \cdot \bm{r}_{j}}{r_{j}^{3}} \,, \\
  \label{eq:21}
  H_{3}(j, k) &\ = a_{e} \frac{\alpha}{2} \Big(i \beta_{j}
                \frac{\bm{\alpha}_{j} \cdot \bm{r}_{jk}}{r_{jk}^{3}} -
                \beta_{j} \bm{\Sigma}_{j} \cdot \frac{\bm{\alpha}_{k} \times
                \bm{r}_{jk}}{r_{jk}^{3}}\Big) \,.
\end{align}
Here, $a_{e} = \alpha / (2\pi) + \ldots$ is the amm of the free electron, and
\[
  \bm{\Sigma} = \Dmatrix{\bm{\sigma}}{0}{0}{\bm{\sigma}} \,.
\]

To the first order in~$1/Z$, the amm correction to the $g$~factor can be
expressed as~\cite{Glazov2004}
\begin{multline}
  \label{eq:22}
  \Delta E_{\mathrm{amm}} = \sum_{c} \Bigg[\delta_{H_1}
  \Big(\bra{vc}I_{\mathrm{Breit}}\ket{vc} -
  \bra{cv}I_{\mathrm{Breit}}\ket{vc}\Big) \\
  + \delta_{H_2} \delta_{V_{\mathrm{mag}}}
  \Big(\bra{vc}I_{\mathrm{Breit}}\ket{vc} -
  \bra{cv}I_{\mathrm{Breit}}\ket{vc}\Big) \\
  + \delta_{V_{\mathrm{mag}}} \Big(\bra{vc}H_{3}\ket{vc} -
  \bra{cv}H_{3}\ket{vc}\Big) \Bigg] \,,
\end{multline}
where $\delta_{V}\left(\ldots\right)$ denotes the first-order perturbation
correction of $\left(\ldots\right)$ induced by $V$ and $I_{\mathrm{Breit}}$ is
the electron-electron interaction operator in Eq.~(\ref{eq:10}) in the Breit
approximation.

The amm approximation is most suitable for light Li-like ions, whereas for
heavy ions the screening-potential approximation becomes preferable.  It should
be mentioned that for B-like ions, the valence electron is in the
$2p_{1/2}$~state and the amm approximation is not applicable at all since it
yields only a part of the $(\Za)^{2}$~contribution, and not a dominant one.

In the present work, we developed a way to combine both the screening-potential
and amm approximations (for Li-like ions).  To this end, we evaluate the amm
correction in Eq.~(\ref{eq:22}) within the screening-potential approximation,
\begin{multline}
  \label{eq:23}
  \Delta E_{\mathrm{amm}}(\mathrm{scr}) = \delta_{H_1}
  \bra{v}V_{\mathrm{scr}}\ket{v} + \delta_{H_2} \delta_{V_{\mathrm{mag}}}
  \bra{v}V_{\mathrm{scr}}\ket{v} \\
  + \delta_{V_{\mathrm{mag}}} \bra{v}H_{3, \mathrm{scr}}\ket{v} \,,
\end{multline}
where
\begin{equation}
  \label{eq:24}
  H_{3,\mathrm{scr}} = -i a_{e} \frac{\alpha}{2} \beta \balpha \cdot
  \bm{\nabla} V_{\mathrm{scr}}(r) \,.
\end{equation}
The difference of Eqs.~(\ref{eq:23}) and (\ref{eq:22}) gives the amm correction
that is \emph{beyond} the screening-potential approximation which can be added
to the results obtained in Section~\ref{sec:scr}.

\subsection{Self-energy results}

\begin{table*}
  \caption{Screened SE~correction to the $g$~factor of the ground state of
    Li-like and B-like ions, in units of $10^{-6}$ (ppm).  ``CH'', ``KS'', and
    ``LDF'' denote results obtained with the core-Hartree, Kohn-Sham, and
    localized Dirac-Fock potentials, respectively.  ``AV'' denotes the averaged
    result.  For Li-like ions, results obtained by two different methods are
    presented, namely, the screening-potential approximation (labelled
    by~``scr'') and the combined screening-potential-and-amm approximation
    (labelled by~``scr+amm'').  For B-like ions, results obtained by the
    screening-potential approximation are listed.}
  \begin{ruledtabular}
    \begin{tabular}{lcdddd}
      \Center{$Z$} & \Center{Method} & \Center{CH}  & \Center{KS}  & \Center{LDF}  & \Center{AV}     \\
      \hline                                                                                         \\[-5pt]
      \multicolumn{3}{l}{Li-like:}                                                                   \\[5pt]
      14           & scr             &  -0.240      &  -0.258      &  -0.257\,(4)  &  -0.257\,(27)   \\
                   & scr+amm         &  -0.250      &  -0.254      &  -0.262\,(4)  &  -0.258\,(17)   \\[2pt]
      18           & scr             &  -0.326      &  -0.349      &  -0.356\,(4)  &  -0.352\,(45)   \\
                   & scr+amm         &  -0.340      &  -0.344      &  -0.363\,(4)  &  -0.354\,(35)   \\[2pt]
      20           & scr             &  -0.371      &  -0.395      &  -0.409\,(4)  &  -0.402\,(58)   \\
                   & scr+amm         &  -0.387      &  -0.390      &  -0.418\,(4)  &  -0.404\,(46)   \\[2pt]
      24           & scr             &  -0.464      &  -0.491      &  -0.523\,(3)  &  -0.507\,(89)   \\
                   & scr+amm         &  -0.485      &  -0.487      &  -0.534\,(3)  &  -0.510\,(72)   \\[2pt]
      32           & scr             &  -0.661      &  -0.695      &  -0.776\,(1)  &  -0.74\,(17)    \\
                   & scr+amm         &  -0.697      &  -0.697      &  -0.792\,(1)  &  -0.75\,(14)    \\[2pt]
      54           & scr             &  -1.318      &  -1.313\,(1) &  -1.689\,(1)  &  -1.50\,(57)    \\
                   & scr+amm         &  -1.429      &  -1.395\,(1) &  -1.718\,(1)  &  -1.56\,(48)    \\[2pt]
      82           & scr             &  -3.007\,(1) &  -2.599\,(1) &  -3.897\,(1)  &  -3.2\,(1.9)    \\
                   & scr+amm         &  -3.249\,(1) &  -3.199\,(1) &  -3.900\,(1)  &  -3.6\,(1.0)    \\[2pt]
      92           & scr             &  -4.168\,(6) &  -3.353\,(6) &  -5.301\,(6)  &  -4.3\,(2.9)    \\
                   & scr+amm         &  -4.384\,(6) &  -4.537\,(6) &  -5.257\,(6)  &  -4.9\,(1.3)    \\
      \multicolumn{3}{l}{B-like:}                                                                    \\[5pt]
      18           & scr             &  -1.042\,(4) &  -0.990\,(4) &  -0.934\,(5)  &  -0.96\,(16)    \\
      20           & scr             &  -1.217\,(4) &  -1.150\,(4) &  -1.093\,(5)  &  -1.12\,(18)    \\
      24           & scr             &  -1.621\,(4) &  -1.528\,(3) &  -1.467\,(5)  &  -1.50\,(23)    \\
      32           & scr             &  -2.670\,(3) &  -2.519\,(1) &  -2.448\,(5)  &  -2.48\,(33)    \\
      54           & scr             &  -7.281\,(1) &  -6.923\,(2) &  -6.870\,(1)  &  -6.90\,(61)    \\
      82           & scr             & -17.374\,(1) & -16.383\,(1) & -16.843\,(1)  & -16.61\,(1.48)  \\
      92           & scr             & -23.030\,(1) & -21.477\,(1) & -22.554\,(1)  & -22.02\,(2.33)  \\
    \end{tabular}
  \end{ruledtabular}
  \label{tab:se_scr}
\end{table*}

Results of our numerical calculations of the screened SE~correction for the
ground state of Li-like and B-like ions are presented in
Table~\ref{tab:se_scr}.  For Li-like ions, we present data obtained with two
approaches: the screening-potential approximation and the combined
screening-potential-and-amm approximation.  With each of the two methods, we
employed three different screening potentials: the core-Hartree (CH), Kohn-Sham
(KS), and localized Dirac-Fock (LDF) potential.  The final result was obtained
as a half-sum of the KS and LDF values, with the error taken as the maximal
difference between the three (CH, KS, and LDF) values, multiplied by a factor
of~$1.5$.  This error estimate is supposed to account for uncalculated effects
that are beyond the screening potential approximation.  We observe that the
combined screening-potential-and-amm approach yields results with a smaller
dependence on the choice of the potential and, as a consequence, to smaller
error bars.  For B-like ions, we present results obtained only with the
screening-potential approximation, since the amm approach is not applicable in
this case.

\begin{table*}
  \caption{Comparison of different calculations of the screened SE~corrections
    to the $g$~factor of the ground state of Li-like ions, in ppm.}
  \begin{ruledtabular}
    \begin{tabular}{ldddd}
      \Center{$Z$} & \Center{This work} & \Center{Full QED}           & \Center{Screening potential} & \Center{AMM}                  \\
                   &                    & \Center{\cite{Volotka2014}} & \Center{\cite{Glazov2006}}   & \Center{\cite{Glazov2004}}    \\
      \hline                                                                                                                         \\[-5pt]
      14           & -0.258\,(17)       & -0.242\,(5)                 &                              & -0.22\,(5)                    \\
      18           & -0.354\,(35)       &                             & -0.24\,(8)                   & -0.29\,(8)                    \\
      20           & -0.404\,(46)       & -0.387\,(7)                 & -0.27\,(10)                  & -0.33\,(10)                   \\
      32           & -0.75\,(14)        &                             & -0.49\,(14)                  & -0.62\,(27)                   \\
      82           & -3.6\,(1.0)        & -3.44\,(2)                  & -3.7\,(1.3)                  & -5.6\,(2.0)                   \\
      92           & -4.9\,(1.3)        & -4.73\,(2)                  & -3.3\,(1.2)                  & -9.2\,(2.6)                   \\
    \end{tabular}
  \end{ruledtabular}
  \label{tab:se_1}
\end{table*}

\begin{table*}
  \caption{SE~corrections to the $g$~factor of the ground state of Li-like and
    B-like ions, in ppm.  Labeling are as follows: ``One-electron (pnt)''
    denotes hydrogenic point-nucleus SE~correction (for the $2s$~state, the
    results are taken from Ref.~\cite{Yerokhin2004}), ``One-electron (fns)''
    denotes the finite nuclear size SE~correction, ``Screening'' denotes the
    screened SE~correction.}
  \begin{ruledtabular}
    \begin{tabular}{ldddd}
      \Center{$Z$} & \Center{One-electron (pnt)} & \Center{One-electron (fns)} & \Center{Screening} & \Center{Total}     \\
      \hline                                                                                                             \\[-5pt]
      \multicolumn{3}{l}{Li-like:}                                                                                       \\[5pt]
      14           & 2324.074\,(3)               &                             &  -0.258\,(17)      &    2323.816\,(18)  \\
      18           & 2325.052\,(5)               &                             &  -0.354\,(35)      &    2324.698\,(35)  \\
      20           & 2325.674\,(5)               &                             &  -0.404\,(46)      &    2325.270\,(47)  \\
      24           & 2327.225\,(5)               &                             &  -0.510\,(72)      &    2326.714\,(73)  \\
      32           & 2331.726\,(6)               & -0.001                      &  -0.75\,(14)       &    2330.98\,(14)   \\
      54           & 2358.184\,(9)               & -0.040                      &  -1.56\,(48)       &    2356.59\,(48)   \\
      82           & 2456.245\,(9)               & -1.540\,(1)                 &  -3.55\,(1.05)     &    2451.16\,(1.05) \\
      92           & 2532.207\,(9)               & -5.488\,(6)                 &  -4.90\,(1.31)     &    2521.82\,(1.31) \\[5pt]
      \multicolumn{3}{l}{B-like:}                                                                                        \\[5pt]
      18           & -768.3723\,(1)              &                             &  -0.96\,(16)       &    -769.34\,(16)   \\
      20           & -766.7594\,(1)              &                             &  -1.12\,(18)       &    -767.88\,(18)   \\
      24           & -762.7517\,(1)              &                             &  -1.50\,(23)       &    -764.25\,(23)   \\
      32           & -751.0481\,(1)              & -0.001                      &  -2.48\,(33)       &    -753.53\,(33)   \\
      54           & -683.2643\,(1)              & -0.012                      &  -6.90\,(62)       &    -690.17\,(62)   \\
      82           & -474.4496\,(4)              & -0.301                      & -16.61\,(1.49)     &    -491.36\,(1.49) \\
      92           & -344.1780\,(3)              & -1.059\,(1)                 & -22.02\,(2.33)     &    -367.25\,(2.33) \\
    \end{tabular}
  \end{ruledtabular}
  \label{tab:se_2}
\end{table*}

Table~\ref{tab:se_1} presents a comparison of the results obtained in this work
for the screened SE~correction for Li-like ions with results of previous
calculations~\cite{Volotka2014, Glazov2006, Glazov2004}.  We observe good
agreement of our data with the results of full QED
calculations~\cite{Volotka2014} and some deviations from the results obtained
by approximate methods.  Our final results for all available SE~corrections to
the $g$~factor of the ground state of Li-like and B-like ions are listed in
Table~\ref{tab:se_2}.

\section{Vacuum polarization}
\label{sec:vacuum_polarization}

\subsection{One-electron vacuum polarization}

\begin{figure}
  \centering
  \subfloat[\label{fig:vp_a}]{\includegraphics[width=0.15\textwidth]{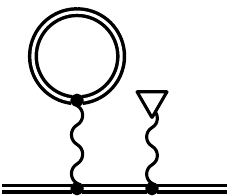}}
  \subfloat[\label{fig:vp_b}]{\includegraphics[width=0.15\textwidth]{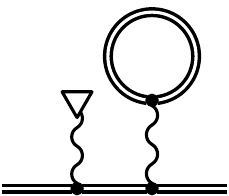}}
  \subfloat[\label{fig:vp_c}]{\includegraphics[width=0.15\textwidth]{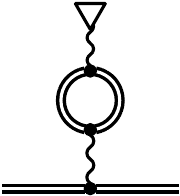}}
  \caption{Feynman diagrams corresponding to the vacuum polarization
    contributions arising from the interaction of the valence electron.}
  \label{fig:vp}
\end{figure}

In the independent electron approximation, i.e. without taking into account the
interactions among the electrons, only the valence electron gives a vacuum
polarization~(VP) contribution to the Zeeman splitting.  The corresponding
diagrams are shown in Fig.~\ref{fig:vp}.  These diagrams are divided into two
groups.  The first group comprises the diagrams in Fig.~\ref{fig:vp_a} and
Fig.~\ref{fig:vp_b}.  They arise due to perturbations of the external wave
functions in the tadpole diagram.  We call this group the electric loop~(EL)
contributions.  The remaining diagram in Fig.~\ref{fig:vp_c} arises due to a
loop correction to the propagator of the photon mitigating the magnetic
interaction.  Accordingly, it is called the magnetic loop~(ML) contribution.
The total VP~contribution to the Zeeman splitting can thus be written as
\begin{equation}
  \label{eq:25}
  \Delta{E}_{\mathrm{VP}} = \Delta{E}_{\mathrm{VP, EL}} +
  \Delta{E}_{\mathrm{VP, ML}}
\end{equation}

For the computation of the EL~contribution, we note that the tadpole part of
the EL~diagrams is equivalent to the insertion of a potential function
$U_{\mathrm{EL}}$ called EL~potential.  A detailed derivation of the formal
expression for $U_{\mathrm{EL}}$ is given in Ref.~\cite{Soff1988}.  It reads
\begin{equation}
  \label{eq:26}
  U_{\mathrm{EL}}(\bm{x}) = \frac{i\alpha}{2\pi} \int d^3{y} \,
  \frac{1}{|\bm{x} - \bm{y}|} \int_{C_{F}} d\omega \, \tr(G(\omega, \bm{y},
  \bm{y})) \,,
\end{equation}
where, again, $G$ denotes the Dirac Coulomb Green's function and $C_{F}$ is the
usual Feynman integration contour.  The contribution to the energy shift is
then
\begin{equation}
  \label{eq:27}
  \Delta{E}_\mathrm{VP, EL} = 2 \bra{v}U_{\mathrm{EL}}\ket{\delta{v}} \,,
\end{equation}
where the first-order perturbation $\ket{\delta{v}}$ of the reference state is
given by Eq.~(\ref{eq:8}).

The expression in Eq.~(\ref{eq:26}) is divergent and needs to be renormalized.
To this end, the potential $U_{\mathrm{EL}}$ is expanded in powers of the
nuclear coupling strength~$Z\alpha$.  This corresponds to an expansion of the
loop in Fig.~\ref{fig:vp_a} and in Fig.~\ref{fig:vp_b} in terms of the free
electron propagator and interactions with the nucleus.  Due to Furry's theorem,
only odd powers of $Z\alpha$ contribute.

The leading term is of order $Z\alpha$ and is called the Uehling contribution.
This term is charge divergent.  After renormalization, it results in a finite
potential called the Uehling potential and is given by~\cite{Fullerton1976}
\begin{equation}
  \label{eq:28}
  U_{\mathrm{Ue}}(\bm{x}) = -\frac{2}{3} \frac{\alpha}{\pi} Z\alpha \int
  d^{3}{y} \, \frac{\varrho(\bm{y})}{|\bm{x} - \bm{y}|} K_{1}(2 |\bm{x} -
  \bm{y}|) \,,
\end{equation}
where $\varrho$ denotes the nuclear charge distribution normalized to one and
where
\begin{equation}
  \label{eq:29}
  K_{1}(x) = \int_{1}^{\infty} dt \, e^{-x t} \left(1 + \frac{1}{2
      t^{2}}\right) \frac{\sqrt{t^{2} - 1}}{t^{2}} \,.
\end{equation}
For our computations of the Uehling potential, we use analytical formulas
resulting from a homogeneously charged sphere as nucleus which have been
derived in Ref.~\cite{Klarsfeld1977}.

The contribution of higher order in $Z\alpha$ to the EL~potential is called the
Wichmann-Kroll potential $U_{\mathrm{WK}}$~\cite{Wichmann1956}.  We use the
expressions in Ref.~\cite{Soff1988} to obtain a partial-wave expansion for the
contributions to the $g$~factor from the partial-wave expansion of the
Wichmann-Kroll potential given by
\begin{equation}
  \label{eq:30}
  U_{\mathrm{WK}}(\bm{x}) = \sum_{|\kappa|=1}^{\infty}
  U_{\mathrm{WK}}^{|\kappa|}(\bm{x}) \,.
\end{equation}
We truncate the partial-wave expansion of the $g$~factor at a finite value of
$|\kappa|$, typically $|\kappa| = 11$, and estimate the remainder by fitting
polynomials in $1/|\kappa|$ to the tail of the partial-wave contributions.  In
Ref.~\cite{Soff1988} the nucleus is taken to be a spherical shell and
analytical solutions for the Dirac-Coulomb Green's function are used in the
calculations.  We, however, use a homogeneously charged sphere as a nucleus
and, thus, compute the Dirac-Coulomb Green's function numerically, much in the
spirit of Refs.~\cite{Yerokhin2011, Yerokhin2013}.  The numerical calculation
is performed using the method of Refs.~\cite{Salvat1991, Salvat1995} for
solving the stationary Dirac equation.  We also use approximate expressions for
this potential derived in Ref.~\cite{Fainshtein1991} for point-like nuclei to
check our numerical calculations.  The total EL potential is then
\begin{equation}
  \label{eq:31}
  U_{\mathrm{EL}} = U_{\mathrm{Ue}} + U_{\mathrm{WK}} \,.
\end{equation}

In the case of the ML~contribution, the effect of the loop can be expressed as
a modification of the vector potential of the external magnetic field.  This
results in a modification of $V_{\mathrm{mag}}$ to $V_{\mathrm{ML}}$ and the
contribution to the Zeeman splitting is given by
\begin{equation}
  \label{eq:32}
  \Delta{E}_{\mathrm{VP, ML}} = \bra{v}V_{\mathrm{ML}}\ket{v} \,,
\end{equation}
where $V_{\mathrm{ML}} = -e \bm{\alpha} \cdot \bm{A}_{\mathrm{ML}}$ and
$\bm{A}_{\mathrm{ML}}$ is the modified vector potential.

In order to compute the modified vector potential, the loop is, again, expanded
in terms of the free-electron propagator and interactions with the nuclear
field.  The leading order term for a point-like nucleus is
$\propto\!(Z\alpha)^{2}$ and has been derived in Refs.~\cite{Lee2005, Lee2007}.
We call this term the Delbr\"{u}ck contribution.  We only take this leading
order term into account and neglect higher order contributions as we expect
them to be small compared to the uncertainties of other contributions to the
$g$~factor.  We obtain
\begin{equation}
  \label{eq:33}
  \bm{A}_{\mathrm{ML}}(\bm{r}) = \bm{A}(\bm{r}) \Pi_{\mathrm{De}}(|\bm{r}|) \,,
\end{equation}
where the polarization function $\Pi_{\mathrm{De}}$ is given by
\begin{equation}
  \label{eq:34}
  \Pi_{\mathrm{De}}(x) = \frac{\alpha}{\pi} (Z\alpha)^{2} \frac{4}{x^{2}}
  \int_{0}^{\infty} dq \, F_{\mathrm{De}}(q) \mathrm{j}_{1}(q x) q x \,.
\end{equation}
In this formula, $\mathrm{j}_{1}$ is the spherical Bessel function of order one
and the function $F_{\mathrm{De}}$ is taken from Ref.~\cite{Lee2007}.
\Change{In Tables~\ref{tab:lilike} and \ref{tab:blike} we estimate the
  uncertainty of the ML~contribution due to higher-order contributions
  conservatively to be $(Z\alpha)^{2} \ln((Z\alpha)^{-2})$ times the
  Delbr\"{u}ck contribution and include this into the calculation of the
  uncertainty of the one-electron VP~contribution.}

\subsection{First-order screened vacuum polarization}

\begin{figure*}
  \centering
  \subfloat[\label{fig:inter_vp_a}]{\includegraphics[width=0.20\textwidth]{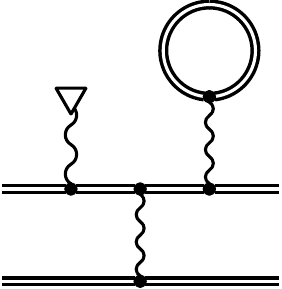}}
  \subfloat[\label{fig:inter_vp_b}]{\includegraphics[width=0.20\textwidth]{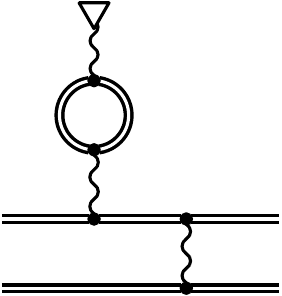}}
  \subfloat[\label{fig:inter_vp_c}]{\includegraphics[width=0.20\textwidth]{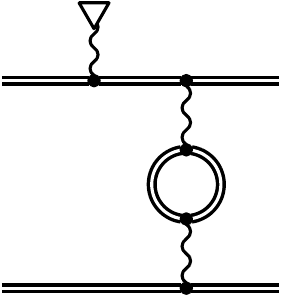}}
  \subfloat[\label{fig:inter_vp_d}]{\includegraphics[width=0.20\textwidth]{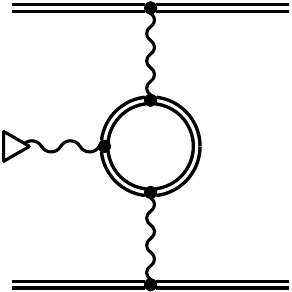}}
  \caption{Vacuum polarization contributions to the two-electron interaction.
    Each of the diagrams shown represents one type of contribution.
    Contributions represented by the first diagram are called electric
    loop~(EL), by the second diagram magnetic loop~(ML), by the third diagram
    electric loop propagator~(ELP), and by the forth magnetic loop
    propagator~(MLP) contributions.}
  \label{fig:inter_vp}
\end{figure*}

Apart from single-electron VP~contributions to the bound-electron $g$~factor,
we calculated the leading order interelectronic-interaction correction to the
VP~effect.  Typical examples of the corresponding Feynman diagrams are depicted
in Fig.~\ref{fig:inter_vp}.  Each of these diagrams represents one of four
groups of contributions.

The first group of contributions is again called electric loop contributions.
The diagram in Fig.~\ref{fig:inter_vp_a} depicts one of the diagrams belonging
to this group.  These EL~contributions arise due to first order perturbative
corrections to the wave functions of the external and the intermediate states
and to the energy levels of the electronic states.  Again, there are reducible
and irreducible contributions to the Zeeman splitting.  Expressions for these
contributions have been derived for Li-like systems in Ref.~\cite{Glazov2010}
using the two-time Green's function formalism~\cite{Shabaev2002}, which can be
readily generalized to the B-like case. 
Alternatively, one can start from Eq.~(\ref{eq:9}) and Eq.~(\ref{eq:11}) and
systematically consider the perturbations of each of the matrix elements.  This
approach has the additional advantage that it also provides a numerical
algorithm to evaluate the contributions.  We used this second approach to
verify the formulas of Ref.~\cite{Glazov2010} and to compute these
contributions.

The second group of contributions, represented by the diagram in
Fig.~\ref{fig:inter_vp_b} with a loop on the photon mitigating the magnetic
interaction, is correspondingly called the magnetic loop contributions.  To
compute these contributions, we need to substitute the magnetic potential
$V_{\mathrm{mag}}$ in Eq.~(\ref{eq:8}) by the magnetic loop potential
$V_{\mathrm{ML}}$ which arises from the modified vector potential in
Eq.~(\ref{eq:33}).

The third group of contributions, represented by Fig.~\ref{fig:inter_vp_c},
arises from a loop correction to the photon propagator mitigating the
interaction between the electrons.  Accordingly, we call it the electric loop
propagator~(ELP) contributions.  We expand the loop in terms of the free
electron propagator and interactions with the nuclear potential.  We take again
only the leading order term into account, which is just the free-electron loop,
since higher-order contributions are expected to be smaller than the
uncertainties of other contributions.  This modifies the photon interaction
operator~$I$ from Eq.~(\ref{eq:10}) into~\cite{Artemyev1997,Artemyev1999}
\begin{multline}
  \label{eq:35}
  \tilde{I}(\varepsilon, \bm{r}_1, \bm{r}_2) = \frac{2}{3} \frac{\alpha}{\pi}
  \int_{1}^{\infty} dt \, \left(1 + \frac{1}{2t^{2}}\right) \frac{\sqrt{t^{2} -
      1}}{t^{2}} \times \\
  \times I\left(\sqrt{\varepsilon^{2} - (2t)^{2}}, \bm{r}_1, \bm{r}_2\right)
  \,.
\end{multline}
The contribution is then obtained by using $\tilde{I}$ instead of $I$ in
Eq.~(\ref{eq:9}) and Eq.~(\ref{eq:11}).

The forth group of contributions is called magnetic loop propagator~(MLP)
contributions.  The diagram in Fig.~\ref{fig:inter_vp_d} represents one of the
contributing diagrams.  If we expand this diagram in terms of the free electron
propagator and interactions with the nuclear field, then, due to Furry's
theorem, the leading order contribution will have four vertices.  As such, its
leading contribution is of higher order than the ELP~contribution.  Thus, we
neglect these terms anticipating that their contribution will be small.

Higher order interaction-effects (i.e. of order $1/Z^2$ or higher) have been
estimated and given as uncertainty of the first-order screened vacuum
polarization result.  For Li-like ions, this effect has been calculated using a
screening potential approach.  For B-like ions, we expect these terms of higher
order to be too small to be visible compared to the uncertainties of the other
contributions for most of the ions considered in this work.  Thus, we estimate
the uncertainty due to higher-order contributions to be $10\%$ of the
first-order contributions.

\subsection{Vacuum polarization results}

\begin{table*}
  \caption{Single-electron VP corrections to the $g$~factor of the ground state
    of Li-like and B-like ions, in units of ppm.  The uncertainty of the
    Uehling contribution results from the uncertainty of the nuclear RMS taken
    from Ref.~\cite{Angeli2013}.  The uncertainty of the Wichmann-Kroll
    contribution is the combined uncertainty due to the nuclear RMS and the
    extrapolation of the partial-wave series.}
  \begin{ruledtabular}
    \begin{tabular}{lddd}
      \Center{$Z$} & \Center{EL, Uehling}       & \Center{EL, Wichmann-Kroll} & \Center{ML, Delbr\"{u}ck} \\
      \hline                                                                                              \\[-5pt]
      \multicolumn{4}{l}{Li-like:}                                                                        \\[5pt]
      18           &  -0.080\,041\,83\,(2)      & 0.000\,244\,93\,(4)         & 0.001\,102\,6             \\
      20           &  -0.120\,944\,51\,(3)      & 0.000\,448\,74\,(4)         & 0.001\,850\,7             \\
      24           &  -0.247\,384\,1\,(2)       & 0.001\,275\,7\,(1)          & 0.004\,524\,6             \\
      32           &  -0.771\,498\,4\,(4)       & 0.006\,605\,(4)             & 0.018\,438                \\
      54           &  -6.622\,35\,(5)           & 0.134\,88\,(1)              & 0.234\,30                 \\
      82           & -46.814\,5\,(4)            & 1.754\,1(1)                 & 1.796\,6                  \\
      92           & -87.661\,(4)               & 3.796\,7(3)                 & 3.168                     \\[5pt]
      \multicolumn{4}{l}{B-like:}                                                                         \\[5pt]
      18           &  -0.000\,418\,694\,19\,(2) & 0.000\,002\,44\,(3)         & 0.000\,413\,11            \\
      20           &  -0.000\,789\,094\,95\,(4) & 0.000\,005\,49\,(5)         & 0.000\,706\,75            \\
      24           &  -0.002\,372\,519\,(1)     & 0.000\,022\,4\,(1)          & 0.001\,797\,2             \\
      32           &  -0.013\,710\,925\,(2)     & 0.000\,209\,(3)             & 0.007\,950\,5             \\
      54           &  -0.376\,778\,(1)          & 0.012\,93\,(2)              & 0.128\,97                 \\
      82           &  -7.250\,91\,(4)           & 0.444\,58(5)                & 1.410\,8                  \\
      92           & -18.394\,5\,(3)            & 1.289\,7(1)                 & 2.881\,5                  \\
    \end{tabular}
  \end{ruledtabular}
  \label{tab:vp_single}
\end{table*}

We present our results of the single-electron VP~correction for the ground
state of Li-like and B-like ions in Table~\ref{tab:vp_single}.  The
contributions are divided into EL and ML~contributions according to our
discussion above.  The Uehling and Wichmann-Kroll contributions to the EL~term
are listed separately.  Both contributions are calculated taking the nuclear
size into account.  The uncertainties result from the quoted uncertainties of
the RMS~radii in Ref.~\cite{Angeli2013}.  \Change{Compared to the uncertainties
  of the other contributions to the $g$~factor, we expect the dependence on the
  nuclear model of higher-order corrections to be of no relevance.}  In the
case of the Wichmann-Kroll contributions, the uncertainties additionally
include the uncertainties from the truncation of the partial-wave expansion.
The Delbr\"{u}ck contribution has been calculated for a point-like nucleus
using the formulas of Refs.~\cite{Lee2005, Lee2007}.  We observe that the
Uehling terms are the largest contributions in magnitude for both the Li-like
and B-like case.  In the Li-like case, we see that while the Delbr\"{u}ck
contribution is larger than the Wichmann-Kroll contribution for low nuclear
charges, this changes for higher nuclear charges.  We also observe that in the
B-like case the Uehling and Delbr\"{u}ck contributions cancel each other to a
significant degree for low nuclear charges.

In addition, we compared our numerical results for the Uehling VP correction
with the $Z\alpha$ expansion.  In Ref.~\cite{Karshenboim2001}, a $Z\alpha$
expansion formula for the Uehling correction was derived up to order
$(Z\alpha)^{7}$ for the $1s$ state only.  We derived an approximation formula
for the Uehling correction in the $2p_{1/2}$ state using non-relativistic
expansions of both the bound-electron wave functions and the wave functions
perturbed linearly by a magnetic field (see Ref.~\cite{Shabaev2003} for the
derivation of the perturbed wave function).  The formula for the Uehling
correction is
\begin{equation}
  \label{eq:36}
  g_{\mathrm{Ue},2p} = \frac{8}{3} \int \dd{r} U_{\mathrm{Ue}}(r) (G_{2p}(r)
  X_{2p}(r) + F_{2p}(r) Y_{2p}(r)) \,,
\end{equation}
with the radial components of the bound electron wave function $G_{2p}$ and
$F_{2p}$ and the magnetic wave function $X_{2p}$ and $Y_{2p}$.  For the
$2p_{1/2}$~state, we find the following non-relativistic expansions of the wave
functions:
\begin{align}
  G_{2p}(r) \approx & -\left(\frac{Z\alpha}{2}\right)^{\frac{1}{2}}
                      \frac{(Z\alpha r)^{2}}{2\sqrt{3}}
                      \exp\left(-\frac{Z\alpha r}{2}\right) \,, \\ 
  F_{2p}(r) \approx & -\left(\frac{Z\alpha}{2}\right)^{\frac{3}{2}} \frac{3
                      Z\alpha r}{2\sqrt{3}} \exp\left(-\frac{Z\alpha
                      r}{2}\right) \,. 
\end{align}
For the radial components of the magnetic wave function, we find
\begin{align}
  X_{2p}(r) & \approx \frac{3}{2} \, r \, F_{2p}(r) - \frac{1}{2} \, G_{2p}(r)
              \,, \\ 
  Y_{2p}(r) & \approx \frac{1}{2} \, r \, G_{2p}(r) + \frac{3}{2} \, F_{2p}(r)
              \,,
\end{align}
using $E_{2p} = 1 + \mathcal{O}\left((Z\alpha)^2\right)$.  With these
non-relativistic wave functions, the radial integration in Eq.~(\ref{eq:36}) as
well as the remaining integration in the representation of the Uehling
potential~\cite{Peskin1995} were carried out analytically to obtain
\begin{equation}
  \label{eq:37}
  g_{\mathrm{Ue}, 2p} \approx -\frac{31}{840} \frac{\alpha}{\pi} (Z\alpha)^{6}
  \,.
\end{equation}
We would like to point out that one has to employ the exact representation of
the Uehling potential in the derivation of this formula since we would obtain
the incomplete result
\begin{equation}
  \label{eq:38}
  g_{\mathrm{Ue}, 2p, \mathrm{incomplete}} \approx -\frac{1}{40}
  \frac{\alpha}{\pi} (Z\alpha)^{6}
\end{equation}
by using the $\delta$~function approximation of the Uehling potential.  For the
$1s$~state, however, the $\delta$~function approximation of the Uehling
potential is sufficient to derive the well-known lowest-order contribution to
the Uehling correction (e.g. \cite{Karshenboim2001})
\begin{equation}
  \label{eq:39}
  g_{\mathrm{Ue}, 1s} \approx -\frac{16}{15} \frac{\alpha}{\pi} (Z\alpha)^{4}
  \,.
\end{equation}
Our numerical all-order results for the Uehling correction were found to be in
good agreement with the approximation formula in Eq.~(\ref{eq:37}) for low~$Z$.

\begin{table*}
  \caption{First-order VP screening correction to the $g$~factor of the ground
    state of Li-like and B-like ions, in units of ppm.  The uncertainty of the
    Uehling contribution results from the uncertainty of the nuclear RMS taken
    from Ref.~\cite{Angeli2013}.  The uncertainty of the Wichmann-Kroll
    contribution is the combined uncertainty due to the nuclear RMS and the
    extrapolation of the partial-wave series.}
  \begin{ruledtabular}
    \begin{tabular}{ldddd}
      \Center{$Z$} & \Center{EL, Uehling} & \Center{EL, Wichmann-Kroll} & \Center{ELP, Uehling} & \Center{ML, Delbr\"{u}ck} \\
      \hline                                                                                                                \\[-5pt]
      \multicolumn{5}{l}{Li-like:}                                                                                          \\[5pt]
      18           & 0.012\,746\,390(3)   & -0.000\,038\,987\,(5)       & -0.000\,085\,6        & -0.000\,163\,3            \\
      20           & 0.017\,346\,904(4)   & -0.000\,064\,318\,(7)       & -0.000\,118           & -0.000\,246\,8            \\
      24           & 0.029\,614\,54(3)    & -0.000\,152\,52\,(1)        & -0.000\,205           & -0.000\,503\,0            \\
      32           & 0.069\,500\,41(3)    & -0.000\,593\,2\,(3)         & -0.000\,499           & -0.001\,537               \\
      54           & 0.356\,946(3 )       & -0.007\,203\,3(5)           & -0.002\,71            & -0.011\,5                 \\
      82           & 1.676\,36(2)         & -0.061\,650\,(4)            & -0.014                & -0.056                    \\
      92           & 2.800\,2(2)          & -0.118\,67\,(2)             & -0.024                & -0.085                    \\[5pt]
      \multicolumn{5}{l}{B-like:}                                                                                           \\[5pt]
      18           & 0.006\,522(2)        & -0.000\,021\,16\,(1)        & 0.000\,003\,6         & -0.000\,134\,1            \\
      20           & 0.008\,987(2)        & -0.000\,035\,59\,(2)        & 0.000\,011            & -0.000\,207               \\
      24           & 0.015\,766(2)        & -0.000\,088\,00\,(7)        & 0.000\,044            & -0.000\,443               \\
      32           & 0.039\,435(5)        & -0.000\,374\,(2)            & 0.000\,26             & -0.001\,499               \\
      54           & 0.257\,1(2)          & -0.006\,256\,(7)            & 0.005\,0              & -0.015\,5                 \\
      82           & 1.853(2)             & -0.088\,1\,(1)              & 0.059                 & -0.126                    \\
      92           & 3.728(2)             & -0.206\,8\,(3)              & 0.13                  & -0.24                     \\
    \end{tabular}
  \end{ruledtabular}
  \label{tab:vp_screened}
\end{table*}

The results for the first-order screened VP~corrections for the ground state of
Li-like and B-like ions are listed in Table~\ref{tab:vp_screened}.  The
contributions are divided according to the groups of diagrams discussed above.
Compared to the single-electron case, we have additionally the
ELP~contribution.  This has been calculated using the leading free-electron
contribution.

\section{Other effects}

\subsection{Nuclear recoil}

In Furry picture calculations, the nucleus is taken to be a source of a
classical background electric field.  This corresponds to taking the nuclear
mass~$M$ to be infinite.  While this often gives a reasonably accurate first
approximation, one needs to take the finite mass of the nucleus into account
for more precise computations of the bound-electron $g$~factor.  This is done
in a perturbative expansion in the small parameter~$1/M$.

In this paper, we include for completeness results for the nuclear-recoil
effect to order~$1/M$ calculated and tabulated in Ref.~\cite{Shabaev2018} for
Li-like ions, and in Refs.~\cite{Glazov2018, Aleksandrov2018} for B-like ions.
\Change{We note that the calculations of the nuclear recoil effect for B-like
  ions have been improved very recently in Ref.~\cite{Glazov2020}}.  No values were
tabulated for Li-like \Ion{}{Xe}{51+} and B-like \Ion{}{Cr}{19+},
\Ion{}{Ge}{27+} and \Ion{}{Xe}{49+} in the given references.  For these ions,
we obtained values and corresponding uncertainties by fitting functions to the
tabulated values, as explained in the following.

For Li- as well as B-like ions, the nuclear recoil correction to the $g$~factor
is written as the sum of a Breit term $\Delta{g}_{\mathrm{Breit}}$ and a QED
term $\Delta{g}_{\mathrm{QED}}$.  For Li-like ions, the Breit term is
parameterized in Ref.~\cite{Shabaev2018} as
\begin{equation}
  \label{eq:40}
  \Delta{g}_{\mathrm{rec, Breit}} = \frac{(Z\alpha)^{2}}{M} \left[A(Z\alpha) +
    \frac{B(Z\alpha)}{Z} + \frac{C(Z\alpha, Z)}{Z^2} \right] \,,
\end{equation}
where the coefficients $A(Z\alpha)$ and $B(Z\alpha)$ denote contributions of
zeroth and first order in~$1/Z$, respectively, and $C(Z\alpha, Z)$ denotes
contributions of second and higher order in~$1/Z$.  The coefficient
$A(Z\alpha)$ was calculated analytically while $B(Z\alpha)$ and $C(Z\alpha, Z)$
were calculated and tabulated numerically in the given reference.  The QED part
is parameterized as
\begin{equation}
  \label{eq:41}
  \Delta{g}_{\mathrm{rec, QED}} = \frac{1}{M} \frac{(Z\alpha)^{5}}{8}
  P(Z\alpha) \,.
\end{equation}
Interelectronic interaction corrections were included using screening potential
approximations.  To obtain values for \Ion{}{Xe}{51+}, we proceeded as follows:
We calculated $A(Z\alpha)$ using the analytical formula given in
Ref.~\cite{Shabaev2018}.  For $B(Z\alpha)$ and $C(Z\alpha, Z)$, we fitted
polynomials in $Z\alpha$ to the tabulated values in the reference.  We use the
$a + b\, (Z\alpha)^{2} + c\, (Z\alpha)^{4}$ to fit $B(Z\alpha)$ and
$a + b\, (Z\alpha) + c\, (Z\alpha)^{2} + d\, (Z\alpha)^{3}$ to fit
$Z\alpha\, C(Z\alpha, Z)$.  For the QED part, given by $Z\alpha\,P(Z\alpha)$,
we used the function
$a\, \ln(Z\alpha) + b + c\, (Z\alpha) + d\, (Z\alpha)^{5}$.

For B-like ions, the Breit term is parameterized in Refs.~\cite{Glazov2018,
  Aleksandrov2018} as
\begin{equation}
  \label{eq:42}
  \Delta{g}_{\mathrm{rec, Breit}} = \frac{1}{M} \left[A_{\mathrm{L}}(Z\alpha) + 
    \frac{B(Z\alpha)}{Z}\right] \,,
\end{equation}
where the coefficients $A_{\mathrm{L}}(Z\alpha)$ and $B(Z\alpha)$ again denote
contributions of zeroth and first order in~$1/Z$, respectively.  The
subscript~L on the coefficient denotes that only lower-order terms in $Z\alpha$
are included.  The QED part is given as
\begin{equation}
  \label{eq:43}
  \Delta{g}_{\mathrm{rec, QED}} = \frac{1}{M}
  \left[A^{2\mathrm{el}}_{\mathrm{H}}(Z\alpha) + \frac{(Z\alpha)^{3}}{8}
    P(Z\alpha)\right] \,,
\end{equation}
where the subscript~H denotes higher-order terms in $Z\alpha$ and the
superscript~2el denotes two-electron contributions to the recoil correction.
Again, interelectronic interactions are included using screening potential
approximations.  To obtain the recoil correction for \Ion{}{Ca}{15+},
\Ion{}{Cr}{19+}, and \Ion{}{Xe}{49+}, we fit the function
$a + b\, (Z\alpha)^{2} + c\, (Z\alpha)^{3} + d\, (Z\alpha)^{7}$ to the data for
$Z\alpha\,[A_{\mathrm{L}}(Z\alpha) + B(Z\alpha)/Z]$ tabulated in
Refs.~\cite{Glazov2018, Aleksandrov2018}, the function
$a + b\, (Z\alpha)^{2} + c\, (Z\alpha)^{4} + d\, (Z\alpha)^{6}$ to the values
for $A^{2\mathrm{el}}_{\mathrm{H}}(Z\alpha)$, and the function
$a + b\, Z\alpha + c\, (Z\alpha)^{2}$ to the data for $Z\alpha\,P(Z\alpha)$ for
small values of $Z$ tabulated in Ref.~\cite{Aleksandrov2018}.

\subsection{Two-loop effects}

For the calculation of two-loop contributions in the independent electron
approximation, we use formulas from Refs.~\cite{Shabaev2002-1, Grotch1973}.
These formulas assume a point-like nucleus and are perturbative in the
nuclear-coupling strength~$Z\alpha$.

In the Li-like case, the formula includes terms up to order
order~$(Z\alpha)^{2}$ and reads~\cite{Shabaev2002-1}
\begin{equation}
  \label{eq:44}
  g_{2s, \mathrm{two-loop}} = -2 \left(1 + \frac{1}{24} (Z\alpha)^{2}\right)
  C_{4} \left(\frac{\alpha}{\pi}\right)^{2} \,,
\end{equation}
where $C_{4}$ denotes the coefficient of the $(\alpha/\pi)^{2}$ contribution in
the expansion of the free-electron magnetic anomaly~$a_{\mathrm{e}}$.  The
expansion coefficient is taken from Ref.~\cite{Mohr2016}.  Uncertainties from
higher-order contributions are estimated by using the formula for the
$(Z\alpha)^{4}$ contribution from Ref.~\cite{Pachucki2005}.

For B-like systems, we have an analytic expression to order~$(Z\alpha)^{0}$.
It is given by~\cite{Grotch1973}
\begin{equation}
  \label{eq:45}
  g_{2p_{1/2}, \mathrm{two-loop}} = -\frac{2}{3} C_{4}
  \left(\frac{\alpha}{\pi}\right)^{2} \,.
\end{equation}
We estimate the uncertainty due to terms of higher order in $Z\alpha$ following
the method of Ref.~\cite{Pachucki2005} as
\begin{equation}
  \label{eq:46}
  g_{\mathrm{h.o.}}^{(2)} = 2 g_{\mathrm{h.o.}}^{(1)}
  \frac{g^{(2)}[(Z\alpha)^{0}]}{g^{(1)}[(Z\alpha)^{0}]} \,,
\end{equation}
where $g_{\mathrm{h.o.}}^{(n)}$ is the $n$-loop higher-order QED contribution
and $g^{(n)}[(Z\alpha)^{0}]$ is the $n$-loop $(Z\alpha)^{0}$ QED contribution.
The contribution $g_{\mathrm{h.o.}}^{(1)}$ as well as the contributions of
order $(Z\alpha)^{0}$ are calculated with the formulas of
Ref.~\cite{Grotch1973}.

\section{Results and Summary}
\label{sec:results}

\begin{table*}
  \caption{Contributions to the bound-electron $g$~factor of lithium-like ions.
    The uncertainties given in parentheses indicate the uncertainty of the last
    digit(s).  If no uncertainty is given, all digits of the quoted value are
    significant.}
  \begin{ruledtabular}
    \begin{tabular}{lllll}
      Contribution                           & \Center{\Ion{40}{Ar}{15+}}  & \Center{\Ion{40}{Ca}{17+}}  & \Center{\Ion{52}{Cr}{21+}}  & \Center{\Ion{74}{Ge}{29+}} \\
      \hline
      Dirac value                            & $\Pos{1.997\,108\,8}$       & $\Pos{1.996\,426\,0}$       & $\Pos{1.994\,838\,1}$       & $\Pos{1.990\,752\,3}$      \\
      Finite nuclear size                    & $\Pos{0.000\,000\,0}$       & $\Pos{0.000\,000\,0}$       & $\Pos{0.000\,000\,0}$       & $\Pos{0.000\,000\,2}$      \\
      Electron correlation:                  &                             &                             &                             &                            \\
      \quad one-photon exchange, $(1/Z)^{1}$ & $\Pos{0.000\,414\,5}$       & $\Pos{0.000\,461\,1}$       & $\Pos{0.000\,555\,2}$       & $\Pos{0.000\,746\,5}$      \\
      \quad $(1/Z)^{2+}$, CI-DFS             & $\Neg{0.000\,006\,7(2)}$    & $\Neg{0.000\,006\,7(2)}$    & $\Neg{0.000\,006\,7(2)}$    & $\Neg{0.000\,006\,7(3)}$   \\
      Nuclear recoil                         & $\Pos{0.000\,000\,1}$       & $\Pos{0.000\,000\,1}$       & $\Pos{0.000\,000\,1}$       & $\Pos{0.000\,000\,1}$      \\
      One-loop QED:                          &                             &                             &                             &                            \\
      \quad SE, $(1/Z)^{0}$                  & $\Pos{0.002\,325\,1}$       & $\Pos{0.002\,325\,7}$       & $\Pos{0.002\,327\,2}$       & $\Pos{0.002\,331\,7}$      \\
      \quad SE, $(1/Z)^{1+}$                 & $\Neg{0.000\,000\,4(1)}$    & $\Neg{0.000\,000\,4(1)}$    & $\Neg{0.000\,000\,5(1)}$    & $\Neg{0.000\,000\,8(2)}$   \\
      \quad VP, $(1/Z)^{0}$                  & $\Neg{0.000\,000\,1}$       & $\Neg{0.000\,000\,1}$       & $\Neg{0.000\,000\,2}$       & $\Neg{0.000\,000\,8}$      \\
      \quad VP, $(1/Z)^{1+}$                 & $\Pos{0.000\,000\,0}$       & $\Pos{0.000\,000\,0}$       & $\Pos{0.000\,000\,0}$       & $\Pos{0.000\,000\,1}$      \\
      Two-loop QED                           & $\Neg{0.000\,003\,5}$       & $\Neg{0.000\,003\,5}$       & $\Neg{0.000\,003\,5}$       & $\Neg{0.000\,003\,6(2)}$   \\
      Total theory                           & $\Pos{1.999\,837\,8(2)}$    & $\Pos{1.999\,202\,2(2)}$    & $\Pos{1.997\,709\,7(2)}$    & $\Pos{1.993\,819\,0(4)}$   \\
      Theory, Ref.~\cite{Glazov2004}         & $\Pos{1.999\,837\,75(14)}$  & $\Pos{1.999\,202\,24(17)}$  & $\Pos{1.997\,709\,70(26)}$  & $\Pos{1.993\,819\,14(46)}$ \\
      Theory, Ref.~\cite{Volotka2014}        &                             & $\Pos{1.999\,202\,041(13)}$ &                             &                            \\
      \hline
      Contribution                           & \Center{\Ion{132}{Xe}{51+}} & \Center{\Ion{208}{Pb}{79+}} & \Center{\Ion{238}{U}{89+}}  &                            \\
      \hline \\[-9pt]
      Dirac value                            & $\Pos{1.972\,750\,2}$       & $\Pos{1.932\,002\,9}$       & $\Pos{1.910\,723}$          &                            \\
      Finite nuclear size                    & $\Pos{0.000\,003\,4}$       & $\Pos{0.000\,078\,7(1)}$    & $\Pos{0.000\,242}$          &                            \\
      Electron correlation:                  &                             &                             &                             &                            \\
      \quad one-photon exchange, $(1/Z)^{1}$ & $\Pos{0.001\,306\,2}$       & $\Pos{0.002\,148\,3}$       & $\Pos{0.002\,510}$          &                            \\
      \quad $(1/Z)^{2+}$, CI-DFS             & $\Neg{0.000\,006\,8(3)}$    & $\Neg{0.000\,007\,6(4)}$    & $\Neg{0.000\,008(1)}$       &                            \\
      Nuclear recoil                         & $\Pos{0.000\,000\,2}$       & $\Pos{0.000\,000\,4}$       & $\Pos{0.000\,001}$          &                            \\
      One-loop QED:                          &                             &                             &                             &                            \\
      \quad SE, $(1/Z)^{0}$                  & $\Pos{0.002\,358\,1(1)}$    & $\Pos{0.002\,454\,7}$       & $\Pos{0.002\,527}$          &                            \\
      \quad SE, $(1/Z)^{1+}$                 & $\Neg{0.000\,001\,6(5)}$    & $\Neg{0.000\,003\,6(11)}$   & $\Neg{0.000\,005(1)}$       &                            \\
      \quad VP, $(1/Z)^{0}$                  & $\Neg{0.000\,006\,3(1)}$    & $\Neg{0.000\,043\,2\Change{(7)}}$    & $\Neg{0.000\,081\Change{(1)}}$       &                            \\
      \quad VP, $(1/Z)^{1+}$                 & $\Pos{0.000\,000\,3(1)}$    & $\Pos{0.000\,001\,6(1)}$    & $\Pos{0.000\,003(1)}$       &                            \\
      Two-loop QED                           & $\Neg{0.000\,003\,6(2)}$    & $\Neg{0.000\,003\,6(12)}$   & $\Neg{0.000\,004(2)}$       &                            \\
      Total theory                           & $\Pos{1.976\,400\,1(6)}$    & $\Pos{1.936\,628\,6(18)}$   & $\Pos{1.915\,908(3)}$       &                            \\
      Theory, Ref.~\cite{Glazov2004}         & $\Pos{1.976\,399\,9(14)}$   & $\Pos{1.936\,625\,3(35)}$   & $\Pos{1.915\,900\,2(50)}$   &                            \\
      Theory, Ref.~\cite{Volotka2014}        &                             & $\Pos{1.936\,627\,2(6)}$    & $\Pos{1.915\,904\,8(11)}$   &                            \\
    \end{tabular}
  \end{ruledtabular}
  \label{tab:lilike}
\end{table*}

\begin{table*}
  \caption{Contributions to the bound-electron $g$~factor of boron-like ions.
    \Change{The uncertainties given in parentheses indicate the uncertainty of
      the last digit(s).  If no uncertainty is given, all digits of the quoted
      value are significant.}}
  \begin{ruledtabular}
    \begin{tabular}{lllll}
      Contribution                           & \Center{\Ion{40}{Ar}{13+}}  & \Center{\Ion{40}{Ca}{15+}}  & \Center{\Ion{52}{Cr}{19+}} & \Center{\Ion{74}{Ge}{27+}}  \\
      \hline
      Dirac value                            & $\Pos{0.663\,775\,5}$       & $\Pos{0.663\,092\,7}$       & $\Pos{0.661\,504\,7}$      & $\Pos{0.657\,419\,0}$       \\
      Finite nuclear size                    & $\Pos{0.000\,000\,0}$       & $\Pos{0.000\,000\,0}$       & $\Pos{0.000\,000\,0}$      & $\Pos{0.000\,000\,0}$       \\
      Electron correlation:                  &                             &                             &                            &                             \\
      \quad one-photon exchange, $(1/Z)^{1}$ & $\Pos{0.000\,657\,5}$       & $\Pos{0.000\,732\,0}$       & $\Pos{0.000\,882\,4}$      & $\Pos{0.001\,190\,3}$       \\
      \quad $(1/Z)^{2+}$, CI-DFS             & $\Neg{0.000\,007\,6(4)}$    & $\Neg{0.000\,007\,7(4)}$    & $\Neg{0.000\,008\,2(5)}$   & $\Neg{0.000\,011\,2(7)}$    \\
      Nuclear recoil                         & $\Neg{0.000\,009\,1(2)}$    & $\Neg{0.000\,009\,3(2)}$    & $\Neg{0.000\,007\,3}$      & $\Neg{0.000\,005\,3}$       \\
      One-loop QED:                          &                             &                             &                            &                             \\
      \quad SE, $(1/Z)^{0}$                  & $\Neg{0.000\,768\,4}$       & $\Neg{0.000\,766\,8}$       & $\Neg{0.000\,762\,8}$      & $\Neg{0.000\,751\,1}$       \\
      \quad SE, $(1/Z)^{1+}$                 & $\Neg{0.000\,001\,0(2)}$    & $\Neg{0.000\,001\,1(2)}$    & $\Neg{0.000\,001\,5(2)}$   & $\Neg{0.000\,002\,5(3)}$    \\
      \quad VP, $(1/Z)^{0}$                  & $\Neg{0.000\,000\,0}$       & $\Neg{0.000\,000\,0}$       & $\Neg{0.000\,000\,0}$      & $\Neg{0.000\,000\,0}$       \\
      \quad VP, $(1/Z)^{1+}$                 & $\Pos{0.000\,000\,0}$       & $\Pos{0.000\,000\,0}$       & $\Pos{0.000\,000\,0}$      & $\Pos{0.000\,000\,0}$       \\
      Two-loop QED                           & $\Pos{0.000\,001\,2(1)}$    & $\Pos{0.000\,001\,2(1)}$    & $\Pos{0.000\,001\,2(1)}$   & $\Pos{0.000\,001\,2(1)}$    \\
      Total theory                           & $\Pos{0.663\,648\,1(5)}$    & $\Pos{0.663\,041\,0(5)}$    & $\Pos{0.661\,608\,5(5)}$   & $\Pos{0.657\,840\,4(8)}$    \\
      \Change{Theory, Ref.~\cite{Agababaev2018}}   & \Change{$\Pos{0.663\,648\,8(12)}$} & \Change{$\Pos{0.663\,041\,8(12)}$} & & \\
      \Change{Theory, Ref.~\cite{Shchepetnov2015}} & \Change{$\Pos{0.663\,647\,7(7)}$}  & & & \\
      \Change{Theory, Ref.~\cite{Marques2016}}     & \Change{$\Pos{0.663\,899(2)}$}     & \Change{$\Pos{0.663\,325(56)}$}    & \Change{$\Pos{0.661\,955(68)}$} & \Change{$\Pos{0.658\,314(93)}$} \\
      \Change{Theory, Ref.~\cite{Verdebout2014}}   & \Change{$\Pos{0.663\,728}$}        & \Change{$\Pos{0.663\,130}$}        & \Change{$\Pos{0.661\,714}$}     & \\
      \hline \\[-9pt]
      Contribution                           & \Center{\Ion{132}{Xe}{49+}} & \Center{\Ion{208}{Pb}{77+}} & \Center{\Ion{238}{U}{87+}} &                             \\
      \hline
      Dirac value                            & $\Pos{0.639\,416\,9}$       & $\Pos{0.598\,669\,6}$       & $\Pos{0.577\,389}$         &                             \\
      Finite nuclear size                    & $\Pos{0.000\,000\,1}$       & $\Pos{0.000\,006\,8}$       & $\Pos{0.000\,029}$         &                             \\
      Electron correlation:                  &                             &                             &                            &                             \\
      \quad one-photon exchange, $(1/Z)^{1}$ & $\Pos{0.002\,118\,2}$       & $\Pos{0.003\,654\,9}$       & $\Pos{0.004\,394}$         &                             \\
      \quad $(1/Z)^{2+}$, CI-DFS             & $\Neg{0.000\,011\,0(5)}$    & $\Neg{0.000\,019\,9(7)}$    & $\Neg{0.000\,023}$         &                             \\
      Nuclear recoil                         & $\Neg{0.000\,003\,0}$       & $\Neg{0.000\,001\,8}$       & $\Neg{0.000\,001}$         &                             \\
      One-loop QED:                          &                             &                             &                            &                             \\
      \quad SE, $(1/Z)^{0}$                  & $\Neg{0.000\,683\,3}$       & $\Neg{0.000\,474\,8}$       & $\Neg{0.000\,345}$         &                             \\
      \quad SE, $(1/Z)^{1+}$                 & $\Neg{0.000\,006\,9(6)}$    & $\Neg{0.000\,016\,6(15)}$   & $\Neg{0.000\,022(2)}$      &                             \\
      \quad VP, $(1/Z)^{0}$                  & $\Neg{0.000\,000\,2(1)}$    & $\Neg{0.000\,005\,5(5)}$    & $\Neg{0.000\,014(1)}$      &                             \\
      \quad VP, $(1/Z)^{1+}$                 & $\Pos{0.000\,000\,2(1)}$    & $\Pos{0.000\,001\,7(2)}$    & $\Pos{0.000\,003(1)}$      &                             \\
      Two-loop QED                           & $\Pos{0.000\,001\,2(1)}$    & $\Pos{0.000\,001\,2(3)}$    & $\Pos{0.000\,001(1)}$      &                             \\
      Total theory                           & $\Pos{0.640\,832\,2(8)}$    & $\Pos{0.601\,815\,6(18)}$   & $\Pos{0.581\,411(3)}$      &                             \\
      \Change{Theory, Ref.~\cite{Marques2016}}     & \Change{$\Pos{0.641\,61(18)}$}     & \Change{$\Pos{0.602\,86(33)}$}    & \Change{$\Pos{0.582\,48(40)}$} & \\
    \end{tabular}
  \end{ruledtabular}
  \label{tab:blike}
\end{table*}

In Table~\ref{tab:lilike} and Table~\ref{tab:blike} we present numerical
results for all the contributions discussed in this work, for Li-like and
B-like ions, respectively.  In Table~\ref{tab:lilike}, we include previous
results for the total $g$~factor of Li-like ions from Refs.~\cite{Glazov2004,
  Volotka2014} for comparison.  Our results independently confirm these
calculations within the given uncertainties.  Our results feature a smaller
uncertainty for high-$Z$ ions than the results of Ref.~\cite{Glazov2004}
from~2004 due to an improved calculation of the screened self-energy
contributions by the combination of the screening potential and the amm
methods.  We also include screened vacuum-polarization contributions which
become visible in the high-$Z$ regime. Ref.~\cite{Volotka2014} treats
many-electron QED effects rigorously by the evaluation of the corresponding
photon exchange QED screening diagrams and two-photon exchange diagrams, and
thus provides the most precise results for the few atomic numbers considered
therein. Our value for the $g$~factor of \Ion{40}{Ca}{17+} agrees also with the
experimental value
\begin{equation*}
  g_{\mathrm{exp}}({}^{40}\mathrm{Ca}^{17+}) = 1.999\,202\,040\,5\,(11)
\end{equation*}
from Ref.~\cite{Koehler2016} within the given uncertainties.

For low atomic numbers, i.e. lower than those considered here, the
nonrelativistic QED approach employing explicitly correlated three-electron
wave functions was found to improve the overall theoretical
uncertainty~\cite{Yerokhin2017-1}.  For Li-like $^{28}$Si$^{11+}$, the most
precise experimental and theoretical $g$~factor values can be found in the very
recent Ref.~\cite{Glazov2019}.

\Change{In Table~\ref{tab:blike}, we include previous results for the total
  $g$~factor of B-like ions from Refs.~\cite{Marques2016, Verdebout2014, Agababaev2018,
    Shchepetnov2015}.  Our results confirm the calculations of
  Refs.~\cite{Agababaev2018, Shchepetnov2015} within the uncertainties.  We
  have improved the uncertainties compared to these works due to an improved
  treatment of the self-energy contributions.}

Recently, the $g$~factor of B-like \Ion{40}{Ar}{13+} was measured by the
ALPHATRAP~experiment~\cite{Sturm2019} at the Max Planck Institute for Nuclear
Physics~\cite{Arapoglou2019}.  The experiment constitutes the first
high-precision measurement of the bound-electron $g$~factor of a B-like ion,
greatly improving the previous experimental value~\cite{Soriaorts2007}.
The measured value for \Ion{40}{Ar}{13+} is~\cite{Arapoglou2019}
\begin{equation*}
  g_{\mathrm{exp}}({}^{40}\mathrm{Ar}^{13+}) = 0.663\,648\,455\,32\,(93) \,.
\end{equation*}
Within the given uncertainty, this value agrees with our result listed in
Table~\ref{tab:blike}.  Our calculation in this work agrees also with a combined
theoretical value $g({}^{40}\mathrm{Ar}^{13+}) = 0.663\,648\,12\,(58)$ of
Ref.~\cite{Arapoglou2019}.

Currently, the main limitation for the calculation of the $g$~factor of B-like
argon stems from the contribution resulting from the higher-order
interelectronic interactions.  For heavy ions (from \Ion{}{Xe}{49+} on), the
uncertainties of the screened self-energy contribution dominate over the other
uncertainties.  Also, vacuum-polarization effects become more visible and need
to be taken into account.  Accordingly, the results for the $g$~factors of
B-like ions can be improved by calculating two-photon exchange contributions
(as has been done in Ref.~\cite{Volotka2014} for Li-like ions) and by
rigorously calculating the screened self-energy effects (as has been done in
Refs.~\cite{Glazov2010, Volotka2014} for Li-like ions).  Furthermore, for a
significant increase of the theoretical precision in case of the heaviest
elements such as \Ion{}{Pb}{77+} and \Ion{}{U}{87+}, which are relevant for an
improved determination of the fine-structure constant~$\alpha$, two-loop
contributions need to be calculated non-perturbatively in the nuclear-coupling
strenght~$Z\alpha$.  First milestones have been achieved for the $1s$ ground
state of hydrogen-like systems in Refs.~\cite{Yerokhin2013-1, Sikora2020};
these calculations need to be extended to the $2p$ valence electron of B-like
ions.

In summary, we performed a systematic calculation of interelectronic and
radiative effects on the one-loop level to the ground-state $g$~factor of
Li-like and B-like ions. \Change{These calculations for B-like ions have been
  extended, for the first time, to heavy elements.} The interelectronic
interaction on the level of one-photon exchange has been calculated using
perturbation theory. Many-electron SE~effects have been taken into account
through screening potentials while the leading-order screening effect for
VP~corrections has been calculated explicitly using perturbation
theory. Estimated theoretical uncertainties have been supplemented for each
value.

\section*{Acknowledgements}

This article comprises parts of the PhD thesis work of H.~C., to be submitted
to Heidelberg University, Germany.  This work is part of and supported by the
German Research Foundation (DFG) Collaborative Research Centre ``SFB 1225
(ISOQUANT).''  V.~A.~Y. acknowledges support by the Ministry of Education and
Science of the Russian Federation Grant No.~3.5397.2017/6.7.


%


\begin{thebibliography}{77}%
\makeatletter
\providecommand \@ifxundefined [1]{%
 \@ifx{#1\undefined}
}%
\providecommand \@ifnum [1]{%
 \ifnum #1\expandafter \@firstoftwo
 \else \expandafter \@secondoftwo
 \fi
}%
\providecommand \@ifx [1]{%
 \ifx #1\expandafter \@firstoftwo
 \else \expandafter \@secondoftwo
 \fi
}%
\providecommand \natexlab [1]{#1}%
\providecommand \enquote  [1]{``#1''}%
\providecommand \bibnamefont  [1]{#1}%
\providecommand \bibfnamefont [1]{#1}%
\providecommand \citenamefont [1]{#1}%
\providecommand \href@noop [0]{\@secondoftwo}%
\providecommand \href [0]{\begingroup \@sanitize@url \@href}%
\providecommand \@href[1]{\@@startlink{#1}\@@href}%
\providecommand \@@href[1]{\endgroup#1\@@endlink}%
\providecommand \@sanitize@url [0]{\catcode `\\12\catcode `\$12\catcode
  `\&12\catcode `\#12\catcode `\^12\catcode `\_12\catcode `\%12\relax}%
\providecommand \@@startlink[1]{}%
\providecommand \@@endlink[0]{}%
\providecommand \url  [0]{\begingroup\@sanitize@url \@url }%
\providecommand \@url [1]{\endgroup\@href {#1}{\urlprefix }}%
\providecommand \urlprefix  [0]{URL }%
\providecommand \Eprint [0]{\href }%
\providecommand \doibase [0]{http://dx.doi.org/}%
\providecommand \selectlanguage [0]{\@gobble}%
\providecommand \bibinfo  [0]{\@secondoftwo}%
\providecommand \bibfield  [0]{\@secondoftwo}%
\providecommand \translation [1]{[#1]}%
\providecommand \BibitemOpen [0]{}%
\providecommand \bibitemStop [0]{}%
\providecommand \bibitemNoStop [0]{.\EOS\space}%
\providecommand \EOS [0]{\spacefactor3000\relax}%
\providecommand \BibitemShut  [1]{\csname bibitem#1\endcsname}%
\let\auto@bib@innerbib\@empty
\bibitem [{\citenamefont {Sturm}\ \emph {et~al.}(2011)\citenamefont {Sturm},
  \citenamefont {Wagner}, \citenamefont {Schabinger}, \citenamefont {Zatorski},
  \citenamefont {Harman}, \citenamefont {Quint}, \citenamefont {Werth},
  \citenamefont {Keitel},\ and\ \citenamefont {Blaum}}]{Sturm2011}%
  \BibitemOpen
  \bibfield  {author} {\bibinfo {author} {\bibfnamefont {S.}~\bibnamefont
  {Sturm}}, \bibinfo {author} {\bibfnamefont {A.}~\bibnamefont {Wagner}},
  \bibinfo {author} {\bibfnamefont {B.}~\bibnamefont {Schabinger}}, \bibinfo
  {author} {\bibfnamefont {J.}~\bibnamefont {Zatorski}}, \bibinfo {author}
  {\bibfnamefont {Z.}~\bibnamefont {Harman}}, \bibinfo {author} {\bibfnamefont
  {W.}~\bibnamefont {Quint}}, \bibinfo {author} {\bibfnamefont
  {G.}~\bibnamefont {Werth}}, \bibinfo {author} {\bibfnamefont {C.~H.}\
  \bibnamefont {Keitel}}, \ and\ \bibinfo {author} {\bibfnamefont
  {K.}~\bibnamefont {Blaum}},\ }\href {\doibase 10.1103/PhysRevLett.107.023002}
  {\bibfield  {journal} {\bibinfo  {journal} {Phys. Rev. Lett.}\ }\textbf
  {\bibinfo {volume} {107}},\ \bibinfo {pages} {023002} (\bibinfo {year}
  {2011})}\BibitemShut {NoStop}%
\bibitem [{\citenamefont {Sturm}\ \emph
  {et~al.}(2013{\natexlab{a}})\citenamefont {Sturm}, \citenamefont {Wagner},
  \citenamefont {Kretzschmar}, \citenamefont {Quint}, \citenamefont {Werth},\
  and\ \citenamefont {Blaum}}]{Sturm2013}%
  \BibitemOpen
  \bibfield  {author} {\bibinfo {author} {\bibfnamefont {S.}~\bibnamefont
  {Sturm}}, \bibinfo {author} {\bibfnamefont {A.}~\bibnamefont {Wagner}},
  \bibinfo {author} {\bibfnamefont {M.}~\bibnamefont {Kretzschmar}}, \bibinfo
  {author} {\bibfnamefont {W.}~\bibnamefont {Quint}}, \bibinfo {author}
  {\bibfnamefont {G.}~\bibnamefont {Werth}}, \ and\ \bibinfo {author}
  {\bibfnamefont {K.}~\bibnamefont {Blaum}},\ }\href {\doibase
  10.1103/PhysRevA.87.030501} {\bibfield  {journal} {\bibinfo  {journal} {Phys.
  Rev. A}\ }\textbf {\bibinfo {volume} {87}},\ \bibinfo {pages} {030501}
  (\bibinfo {year} {2013}{\natexlab{a}})}\BibitemShut {NoStop}%
\bibitem [{\citenamefont {Pachucki}\ \emph {et~al.}(2004)\citenamefont
  {Pachucki}, \citenamefont {Jentschura},\ and\ \citenamefont
  {Yerokhin}}]{Pachucki2004}%
  \BibitemOpen
  \bibfield  {author} {\bibinfo {author} {\bibfnamefont {K.}~\bibnamefont
  {Pachucki}}, \bibinfo {author} {\bibfnamefont {U.~D.}\ \bibnamefont
  {Jentschura}}, \ and\ \bibinfo {author} {\bibfnamefont {V.~A.}\ \bibnamefont
  {Yerokhin}},\ }\href {\doibase 10.1103/PhysRevLett.93.150401} {\bibfield
  {journal} {\bibinfo  {journal} {Phys. Rev. Lett.}\ }\textbf {\bibinfo
  {volume} {93}},\ \bibinfo {pages} {150401} (\bibinfo {year}
  {2004})}\BibitemShut {NoStop}%
\bibitem [{\citenamefont {Pachucki}\ \emph {et~al.}(2005)\citenamefont
  {Pachucki}, \citenamefont {Czarnecki}, \citenamefont {Jentschura},\ and\
  \citenamefont {Yerokhin}}]{Pachucki2005}%
  \BibitemOpen
  \bibfield  {author} {\bibinfo {author} {\bibfnamefont {K.}~\bibnamefont
  {Pachucki}}, \bibinfo {author} {\bibfnamefont {A.}~\bibnamefont {Czarnecki}},
  \bibinfo {author} {\bibfnamefont {U.~D.}\ \bibnamefont {Jentschura}}, \ and\
  \bibinfo {author} {\bibfnamefont {V.~A.}\ \bibnamefont {Yerokhin}},\ }\href
  {\doibase 10.1103/PhysRevA.72.022108} {\bibfield  {journal} {\bibinfo
  {journal} {Phys. Rev. A}\ }\textbf {\bibinfo {volume} {72}},\ \bibinfo
  {pages} {022108} (\bibinfo {year} {2005})}\BibitemShut {NoStop}%
\bibitem [{\citenamefont {Karshenboim}\ and\ \citenamefont
  {Milstein}(2002)}]{Karshenboim2002}%
  \BibitemOpen
  \bibfield  {author} {\bibinfo {author} {\bibfnamefont {S.~G.}\ \bibnamefont
  {Karshenboim}}\ and\ \bibinfo {author} {\bibfnamefont {A.~I.}\ \bibnamefont
  {Milstein}},\ }\href {\doibase https://doi.org/10.1016/S0370-2693(02)02930-1}
  {\bibfield  {journal} {\bibinfo  {journal} {Phys. Lett. B}\ }\textbf
  {\bibinfo {volume} {549}},\ \bibinfo {pages} {321 } (\bibinfo {year}
  {2002})}\BibitemShut {NoStop}%
\bibitem [{\citenamefont {Lee}\ \emph {et~al.}(2005)\citenamefont {Lee},
  \citenamefont {Milstein}, \citenamefont {Terekhov},\ and\ \citenamefont
  {Karshenboim}}]{Lee2005}%
  \BibitemOpen
  \bibfield  {author} {\bibinfo {author} {\bibfnamefont {R.~N.}\ \bibnamefont
  {Lee}}, \bibinfo {author} {\bibfnamefont {A.~I.}\ \bibnamefont {Milstein}},
  \bibinfo {author} {\bibfnamefont {I.~S.}\ \bibnamefont {Terekhov}}, \ and\
  \bibinfo {author} {\bibfnamefont {S.~G.}\ \bibnamefont {Karshenboim}},\
  }\href {\doibase 10.1103/PhysRevA.71.052501} {\bibfield  {journal} {\bibinfo
  {journal} {Phys. Rev. A}\ }\textbf {\bibinfo {volume} {71}},\ \bibinfo
  {pages} {052501} (\bibinfo {year} {2005})}\BibitemShut {NoStop}%
\bibitem [{\citenamefont {Yerokhin}\ \emph {et~al.}(2002)\citenamefont
  {Yerokhin}, \citenamefont {Indelicato},\ and\ \citenamefont
  {Shabaev}}]{Yerokhin2002}%
  \BibitemOpen
  \bibfield  {author} {\bibinfo {author} {\bibfnamefont {V.~A.}\ \bibnamefont
  {Yerokhin}}, \bibinfo {author} {\bibfnamefont {P.}~\bibnamefont
  {Indelicato}}, \ and\ \bibinfo {author} {\bibfnamefont {V.~M.}\ \bibnamefont
  {Shabaev}},\ }\href {\doibase 10.1103/PhysRevLett.89.143001} {\bibfield
  {journal} {\bibinfo  {journal} {Phys. Rev. Lett.}\ }\textbf {\bibinfo
  {volume} {89}},\ \bibinfo {pages} {143001} (\bibinfo {year}
  {2002})}\BibitemShut {NoStop}%
\bibitem [{\citenamefont {Yerokhin}\ \emph {et~al.}(2004)\citenamefont
  {Yerokhin}, \citenamefont {Indelicato},\ and\ \citenamefont
  {Shabaev}}]{Yerokhin2004}%
  \BibitemOpen
  \bibfield  {author} {\bibinfo {author} {\bibfnamefont {V.~A.}\ \bibnamefont
  {Yerokhin}}, \bibinfo {author} {\bibfnamefont {P.}~\bibnamefont
  {Indelicato}}, \ and\ \bibinfo {author} {\bibfnamefont {V.~M.}\ \bibnamefont
  {Shabaev}},\ }\href {\doibase 10.1103/PhysRevA.69.052503} {\bibfield
  {journal} {\bibinfo  {journal} {Phys. Rev. A}\ }\textbf {\bibinfo {volume}
  {69}},\ \bibinfo {pages} {052503} (\bibinfo {year} {2004})}\BibitemShut
  {NoStop}%
\bibitem [{\citenamefont {Shabaev}\ and\ \citenamefont
  {Yerokhin}(2002)}]{Shabaev2002-2}%
  \BibitemOpen
  \bibfield  {author} {\bibinfo {author} {\bibfnamefont {V.~M.}\ \bibnamefont
  {Shabaev}}\ and\ \bibinfo {author} {\bibfnamefont {V.~A.}\ \bibnamefont
  {Yerokhin}},\ }\href {\doibase 10.1103/PhysRevLett.88.091801} {\bibfield
  {journal} {\bibinfo  {journal} {Phys. Rev. Lett.}\ }\textbf {\bibinfo
  {volume} {88}},\ \bibinfo {pages} {091801} (\bibinfo {year}
  {2002})}\BibitemShut {NoStop}%
\bibitem [{\citenamefont {Beier}(2000)}]{Beier2000-1}%
  \BibitemOpen
  \bibfield  {author} {\bibinfo {author} {\bibfnamefont {T.}~\bibnamefont
  {Beier}},\ }\href {\doibase https://doi.org/10.1016/S0370-1573(00)00071-5}
  {\bibfield  {journal} {\bibinfo  {journal} {Phys. Rep.}\ }\textbf {\bibinfo
  {volume} {339}},\ \bibinfo {pages} {79 } (\bibinfo {year}
  {2000})}\BibitemShut {NoStop}%
\bibitem [{\citenamefont {Beier}\ \emph {et~al.}(2000)\citenamefont {Beier},
  \citenamefont {Lindgren}, \citenamefont {Persson}, \citenamefont
  {Salomonson}, \citenamefont {Sunnergren}, \citenamefont {H\"affner},\ and\
  \citenamefont {Hermanspahn}}]{Beier2000}%
  \BibitemOpen
  \bibfield  {author} {\bibinfo {author} {\bibfnamefont {T.}~\bibnamefont
  {Beier}}, \bibinfo {author} {\bibfnamefont {I.}~\bibnamefont {Lindgren}},
  \bibinfo {author} {\bibfnamefont {H.}~\bibnamefont {Persson}}, \bibinfo
  {author} {\bibfnamefont {S.}~\bibnamefont {Salomonson}}, \bibinfo {author}
  {\bibfnamefont {P.}~\bibnamefont {Sunnergren}}, \bibinfo {author}
  {\bibfnamefont {H.}~\bibnamefont {H\"affner}}, \ and\ \bibinfo {author}
  {\bibfnamefont {N.}~\bibnamefont {Hermanspahn}},\ }\href {\doibase
  10.1103/PhysRevA.62.032510} {\bibfield  {journal} {\bibinfo  {journal} {Phys.
  Rev. A}\ }\textbf {\bibinfo {volume} {62}},\ \bibinfo {pages} {032510}
  (\bibinfo {year} {2000})}\BibitemShut {NoStop}%
\bibitem [{\citenamefont {Czarnecki}\ \emph {et~al.}(2018)\citenamefont
  {Czarnecki}, \citenamefont {Dowling}, \citenamefont {Piclum},\ and\
  \citenamefont {Szafron}}]{Czarnecki2018}%
  \BibitemOpen
  \bibfield  {author} {\bibinfo {author} {\bibfnamefont {A.}~\bibnamefont
  {Czarnecki}}, \bibinfo {author} {\bibfnamefont {M.}~\bibnamefont {Dowling}},
  \bibinfo {author} {\bibfnamefont {J.}~\bibnamefont {Piclum}}, \ and\ \bibinfo
  {author} {\bibfnamefont {R.}~\bibnamefont {Szafron}},\ }\href {\doibase
  10.1103/PhysRevLett.120.043203} {\bibfield  {journal} {\bibinfo  {journal}
  {Phys. Rev. Lett.}\ }\textbf {\bibinfo {volume} {120}},\ \bibinfo {pages}
  {043203} (\bibinfo {year} {2018})}\BibitemShut {NoStop}%
\bibitem [{\citenamefont {Pachucki}\ and\ \citenamefont
  {Puchalski}(2017)}]{Pachucki2017}%
  \BibitemOpen
  \bibfield  {author} {\bibinfo {author} {\bibfnamefont {K.}~\bibnamefont
  {Pachucki}}\ and\ \bibinfo {author} {\bibfnamefont {M.}~\bibnamefont
  {Puchalski}},\ }\href {\doibase 10.1103/PhysRevA.96.032503} {\bibfield
  {journal} {\bibinfo  {journal} {Phys. Rev. A}\ }\textbf {\bibinfo {volume}
  {96}},\ \bibinfo {pages} {032503} (\bibinfo {year} {2017})}\BibitemShut
  {NoStop}%
\bibitem [{\citenamefont {Yerokhin}\ and\ \citenamefont
  {Harman}(2013)}]{Yerokhin2013-1}%
  \BibitemOpen
  \bibfield  {author} {\bibinfo {author} {\bibfnamefont {V.~A.}\ \bibnamefont
  {Yerokhin}}\ and\ \bibinfo {author} {\bibfnamefont {Z.}~\bibnamefont
  {Harman}},\ }\href {\doibase 10.1103/PhysRevA.88.042502} {\bibfield
  {journal} {\bibinfo  {journal} {Phys. Rev. A}\ }\textbf {\bibinfo {volume}
  {88}},\ \bibinfo {pages} {042502} (\bibinfo {year} {2013})}\BibitemShut
  {NoStop}%
\bibitem [{\citenamefont {Sikora}\ \emph {et~al.}(2020)\citenamefont {Sikora},
  \citenamefont {Yerokhin}, \citenamefont {Oreshkina}, \citenamefont {Cakir},
  \citenamefont {Keitel},\ and\ \citenamefont {Harman}}]{Sikora2020}%
  \BibitemOpen
  \bibfield  {author} {\bibinfo {author} {\bibfnamefont {B.}~\bibnamefont
  {Sikora}}, \bibinfo {author} {\bibfnamefont {V.~A.}\ \bibnamefont
  {Yerokhin}}, \bibinfo {author} {\bibfnamefont {N.~S.}\ \bibnamefont
  {Oreshkina}}, \bibinfo {author} {\bibfnamefont {H.}~\bibnamefont {Cakir}},
  \bibinfo {author} {\bibfnamefont {C.~H.}\ \bibnamefont {Keitel}}, \ and\
  \bibinfo {author} {\bibfnamefont {Z.}~\bibnamefont {Harman}},\ }\href
  {\doibase 10.1103/PhysRevResearch.2.012002} {\bibfield  {journal} {\bibinfo
  {journal} {Phys. Rev. Research}\ }\textbf {\bibinfo {volume} {2}},\ \bibinfo
  {pages} {012002} (\bibinfo {year} {2020})}\BibitemShut {NoStop}%
\bibitem [{\citenamefont {Sturm}\ \emph {et~al.}(2014)\citenamefont {Sturm},
  \citenamefont {K{\"o}hler}, \citenamefont {Zatorski}, \citenamefont {Wagner},
  \citenamefont {Harman}, \citenamefont {Werth}, \citenamefont {Quint},
  \citenamefont {Keitel},\ and\ \citenamefont {Blaum}}]{Sturm2014}%
  \BibitemOpen
  \bibfield  {author} {\bibinfo {author} {\bibfnamefont {S.}~\bibnamefont
  {Sturm}}, \bibinfo {author} {\bibfnamefont {F.}~\bibnamefont {K{\"o}hler}},
  \bibinfo {author} {\bibfnamefont {J.}~\bibnamefont {Zatorski}}, \bibinfo
  {author} {\bibfnamefont {A.}~\bibnamefont {Wagner}}, \bibinfo {author}
  {\bibfnamefont {Z.}~\bibnamefont {Harman}}, \bibinfo {author} {\bibfnamefont
  {G.}~\bibnamefont {Werth}}, \bibinfo {author} {\bibfnamefont
  {W.}~\bibnamefont {Quint}}, \bibinfo {author} {\bibfnamefont {C.~H.}\
  \bibnamefont {Keitel}}, \ and\ \bibinfo {author} {\bibfnamefont
  {K.}~\bibnamefont {Blaum}},\ }\href {\doibase 10.1038/nature13026} {\bibfield
   {journal} {\bibinfo  {journal} {Nature}\ }\textbf {\bibinfo {volume}
  {506}},\ \bibinfo {pages} {467} (\bibinfo {year} {2014})}\BibitemShut
  {NoStop}%
\bibitem [{\citenamefont {K\"{o}hler}\ \emph {et~al.}(2015)\citenamefont
  {K\"{o}hler}, \citenamefont {Sturm}, \citenamefont {Kracke}, \citenamefont
  {Werth}, \citenamefont {Quint},\ and\ \citenamefont {Blaum}}]{Koehler2015}%
  \BibitemOpen
  \bibfield  {author} {\bibinfo {author} {\bibfnamefont {F.}~\bibnamefont
  {K\"{o}hler}}, \bibinfo {author} {\bibfnamefont {S.}~\bibnamefont {Sturm}},
  \bibinfo {author} {\bibfnamefont {A.}~\bibnamefont {Kracke}}, \bibinfo
  {author} {\bibfnamefont {G.}~\bibnamefont {Werth}}, \bibinfo {author}
  {\bibfnamefont {W.}~\bibnamefont {Quint}}, \ and\ \bibinfo {author}
  {\bibfnamefont {K.}~\bibnamefont {Blaum}},\ }\href {\doibase
  10.1088/0953-4075/48/14/144032} {\bibfield  {journal} {\bibinfo  {journal}
  {J. Phys. B}\ }\textbf {\bibinfo {volume} {48}},\ \bibinfo {pages} {144032}
  (\bibinfo {year} {2015})}\BibitemShut {NoStop}%
\bibitem [{\citenamefont {Zatorski}\ \emph {et~al.}(2017)\citenamefont
  {Zatorski}, \citenamefont {Sikora}, \citenamefont {Karshenboim},
  \citenamefont {Sturm}, \citenamefont {K\"ohler-Langes}, \citenamefont
  {Blaum}, \citenamefont {Keitel},\ and\ \citenamefont
  {Harman}}]{Zatorski2017}%
  \BibitemOpen
  \bibfield  {author} {\bibinfo {author} {\bibfnamefont {J.}~\bibnamefont
  {Zatorski}}, \bibinfo {author} {\bibfnamefont {B.}~\bibnamefont {Sikora}},
  \bibinfo {author} {\bibfnamefont {S.~G.}\ \bibnamefont {Karshenboim}},
  \bibinfo {author} {\bibfnamefont {S.}~\bibnamefont {Sturm}}, \bibinfo
  {author} {\bibfnamefont {F.}~\bibnamefont {K\"ohler-Langes}}, \bibinfo
  {author} {\bibfnamefont {K.}~\bibnamefont {Blaum}}, \bibinfo {author}
  {\bibfnamefont {C.~H.}\ \bibnamefont {Keitel}}, \ and\ \bibinfo {author}
  {\bibfnamefont {Z.}~\bibnamefont {Harman}},\ }\href {\doibase
  10.1103/PhysRevA.96.012502} {\bibfield  {journal} {\bibinfo  {journal} {Phys.
  Rev. A}\ }\textbf {\bibinfo {volume} {96}},\ \bibinfo {pages} {012502}
  (\bibinfo {year} {2017})}\BibitemShut {NoStop}%
\bibitem [{\citenamefont {Glazov}\ and\ \citenamefont
  {Shabaev}(2002)}]{Glazov2002}%
  \BibitemOpen
  \bibfield  {author} {\bibinfo {author} {\bibfnamefont {D.~A.}\ \bibnamefont
  {Glazov}}\ and\ \bibinfo {author} {\bibfnamefont {V.~M.}\ \bibnamefont
  {Shabaev}},\ }\href {\doibase https://doi.org/10.1016/S0375-9601(02)00021-X}
  {\bibfield  {journal} {\bibinfo  {journal} {Phys. Lett. A}\ }\textbf
  {\bibinfo {volume} {297}},\ \bibinfo {pages} {408 } (\bibinfo {year}
  {2002})}\BibitemShut {NoStop}%
\bibitem [{\citenamefont {Nefiodov}\ \emph {et~al.}(2002)\citenamefont
  {Nefiodov}, \citenamefont {Plunien},\ and\ \citenamefont
  {Soff}}]{Nefiodov2002}%
  \BibitemOpen
  \bibfield  {author} {\bibinfo {author} {\bibfnamefont {A.~V.}\ \bibnamefont
  {Nefiodov}}, \bibinfo {author} {\bibfnamefont {G.}~\bibnamefont {Plunien}}, \
  and\ \bibinfo {author} {\bibfnamefont {G.}~\bibnamefont {Soff}},\ }\href
  {\doibase 10.1103/PhysRevLett.89.081802} {\bibfield  {journal} {\bibinfo
  {journal} {Phys. Rev. Lett.}\ }\textbf {\bibinfo {volume} {89}},\ \bibinfo
  {pages} {081802} (\bibinfo {year} {2002})}\BibitemShut {NoStop}%
\bibitem [{\citenamefont {Zatorski}\ \emph {et~al.}(2012)\citenamefont
  {Zatorski}, \citenamefont {Oreshkina}, \citenamefont {Keitel},\ and\
  \citenamefont {Harman}}]{Zatorski2012}%
  \BibitemOpen
  \bibfield  {author} {\bibinfo {author} {\bibfnamefont {J.}~\bibnamefont
  {Zatorski}}, \bibinfo {author} {\bibfnamefont {N.~S.}\ \bibnamefont
  {Oreshkina}}, \bibinfo {author} {\bibfnamefont {C.~H.}\ \bibnamefont
  {Keitel}}, \ and\ \bibinfo {author} {\bibfnamefont {Z.}~\bibnamefont
  {Harman}},\ }\href {\doibase 10.1103/PhysRevLett.108.063005} {\bibfield
  {journal} {\bibinfo  {journal} {Phys. Rev. Lett.}\ }\textbf {\bibinfo
  {volume} {108}},\ \bibinfo {pages} {063005} (\bibinfo {year}
  {2012})}\BibitemShut {NoStop}%
\bibitem [{\citenamefont {Sturm}\ \emph
  {et~al.}(2013{\natexlab{b}})\citenamefont {Sturm}, \citenamefont {Werth},\
  and\ \citenamefont {Blaum}}]{Sturm2013-1}%
  \BibitemOpen
  \bibfield  {author} {\bibinfo {author} {\bibfnamefont {S.}~\bibnamefont
  {Sturm}}, \bibinfo {author} {\bibfnamefont {G.}~\bibnamefont {Werth}}, \ and\
  \bibinfo {author} {\bibfnamefont {K.}~\bibnamefont {Blaum}},\ }\href
  {\doibase 10.1002/andp.201300052} {\bibfield  {journal} {\bibinfo  {journal}
  {Ann. Phys. (N. Y.)}\ }\textbf {\bibinfo {volume} {525}},\ \bibinfo {pages}
  {620} (\bibinfo {year} {2013}{\natexlab{b}})}\BibitemShut {NoStop}%
\bibitem [{\citenamefont {Quint}\ \emph {et~al.}(2001)\citenamefont {Quint},
  \citenamefont {Dilling}, \citenamefont {Djekic}, \citenamefont {H\"affner},
  \citenamefont {Hermanspahn}, \citenamefont {Kluge}, \citenamefont {Marx},
  \citenamefont {Moore}, \citenamefont {Rodriguez}, \citenamefont
  {Sch\"onfelder}, \citenamefont {Sikler}, \citenamefont {Valenzuela},
  \citenamefont {Verd\'u}, \citenamefont {Weber},\ and\ \citenamefont
  {Werth}}]{Quint2001}%
  \BibitemOpen
  \bibfield  {author} {\bibinfo {author} {\bibfnamefont {W.}~\bibnamefont
  {Quint}}, \bibinfo {author} {\bibfnamefont {J.}~\bibnamefont {Dilling}},
  \bibinfo {author} {\bibfnamefont {S.}~\bibnamefont {Djekic}}, \bibinfo
  {author} {\bibfnamefont {H.}~\bibnamefont {H\"affner}}, \bibinfo {author}
  {\bibfnamefont {N.}~\bibnamefont {Hermanspahn}}, \bibinfo {author}
  {\bibfnamefont {H.-J.}\ \bibnamefont {Kluge}}, \bibinfo {author}
  {\bibfnamefont {G.}~\bibnamefont {Marx}}, \bibinfo {author} {\bibfnamefont
  {R.}~\bibnamefont {Moore}}, \bibinfo {author} {\bibfnamefont
  {D.}~\bibnamefont {Rodriguez}}, \bibinfo {author} {\bibfnamefont
  {J.}~\bibnamefont {Sch\"onfelder}}, \bibinfo {author} {\bibfnamefont
  {G.}~\bibnamefont {Sikler}}, \bibinfo {author} {\bibfnamefont
  {T.}~\bibnamefont {Valenzuela}}, \bibinfo {author} {\bibfnamefont
  {J.}~\bibnamefont {Verd\'u}}, \bibinfo {author} {\bibfnamefont
  {C.}~\bibnamefont {Weber}}, \ and\ \bibinfo {author} {\bibfnamefont
  {G.}~\bibnamefont {Werth}},\ }\href@noop {} {\bibfield  {journal} {\bibinfo
  {journal} {Hyperfine Interact.}\ }\textbf {\bibinfo {volume} {132}},\
  \bibinfo {pages} {457} (\bibinfo {year} {2001})}\BibitemShut {NoStop}%
\bibitem [{\citenamefont {Kluge}\ \emph {et~al.}(2008)\citenamefont {Kluge},
  \citenamefont {Beier}, \citenamefont {Blaum}, \citenamefont {Dahl},
  \citenamefont {Eliseev}, \citenamefont {Herfurth}, \citenamefont {Hofmann},
  \citenamefont {Kester}, \citenamefont {Koszudowski}, \citenamefont
  {Kozhuharov} \emph {et~al.}}]{Kluge2008}%
  \BibitemOpen
  \bibfield  {author} {\bibinfo {author} {\bibfnamefont {H.-J.}\ \bibnamefont
  {Kluge}}, \bibinfo {author} {\bibfnamefont {T.}~\bibnamefont {Beier}},
  \bibinfo {author} {\bibfnamefont {K.}~\bibnamefont {Blaum}}, \bibinfo
  {author} {\bibfnamefont {L.}~\bibnamefont {Dahl}}, \bibinfo {author}
  {\bibfnamefont {S.}~\bibnamefont {Eliseev}}, \bibinfo {author} {\bibfnamefont
  {F.}~\bibnamefont {Herfurth}}, \bibinfo {author} {\bibfnamefont
  {B.}~\bibnamefont {Hofmann}}, \bibinfo {author} {\bibfnamefont
  {O.}~\bibnamefont {Kester}}, \bibinfo {author} {\bibfnamefont
  {S.}~\bibnamefont {Koszudowski}}, \bibinfo {author} {\bibfnamefont
  {C.}~\bibnamefont {Kozhuharov}},  \emph {et~al.},\ }\href@noop {} {\bibfield
  {journal} {\bibinfo  {journal} {Adv. Quantum Chem.}\ }\textbf {\bibinfo
  {volume} {53}},\ \bibinfo {pages} {83} (\bibinfo {year} {2008})}\BibitemShut
  {NoStop}%
\bibitem [{\citenamefont {Shabaev}\ \emph {et~al.}(2006)\citenamefont
  {Shabaev}, \citenamefont {Glazov}, \citenamefont {Oreshkina}, \citenamefont
  {Volotka}, \citenamefont {Plunien}, \citenamefont {Kluge},\ and\
  \citenamefont {Quint}}]{Shabaev2006}%
  \BibitemOpen
  \bibfield  {author} {\bibinfo {author} {\bibfnamefont {V.~M.}\ \bibnamefont
  {Shabaev}}, \bibinfo {author} {\bibfnamefont {D.~A.}\ \bibnamefont {Glazov}},
  \bibinfo {author} {\bibfnamefont {N.~S.}\ \bibnamefont {Oreshkina}}, \bibinfo
  {author} {\bibfnamefont {A.~V.}\ \bibnamefont {Volotka}}, \bibinfo {author}
  {\bibfnamefont {G.}~\bibnamefont {Plunien}}, \bibinfo {author} {\bibfnamefont
  {H.-J.}\ \bibnamefont {Kluge}}, \ and\ \bibinfo {author} {\bibfnamefont
  {W.}~\bibnamefont {Quint}},\ }\href {\doibase 10.1103/PhysRevLett.96.253002}
  {\bibfield  {journal} {\bibinfo  {journal} {Phys. Rev. Lett.}\ }\textbf
  {\bibinfo {volume} {96}},\ \bibinfo {pages} {253002} (\bibinfo {year}
  {2006})}\BibitemShut {NoStop}%
\bibitem [{\citenamefont {Yerokhin}\ \emph
  {et~al.}(2016{\natexlab{a}})\citenamefont {Yerokhin}, \citenamefont
  {Berseneva}, \citenamefont {Harman}, \citenamefont {Tupitsyn},\ and\
  \citenamefont {Keitel}}]{Yerokhin2016}%
  \BibitemOpen
  \bibfield  {author} {\bibinfo {author} {\bibfnamefont {V.~A.}\ \bibnamefont
  {Yerokhin}}, \bibinfo {author} {\bibfnamefont {E.}~\bibnamefont {Berseneva}},
  \bibinfo {author} {\bibfnamefont {Z.}~\bibnamefont {Harman}}, \bibinfo
  {author} {\bibfnamefont {I.~I.}\ \bibnamefont {Tupitsyn}}, \ and\ \bibinfo
  {author} {\bibfnamefont {C.~H.}\ \bibnamefont {Keitel}},\ }\href {\doibase
  10.1103/PhysRevLett.116.100801} {\bibfield  {journal} {\bibinfo  {journal}
  {Phys. Rev. Lett.}\ }\textbf {\bibinfo {volume} {116}},\ \bibinfo {pages}
  {100801} (\bibinfo {year} {2016}{\natexlab{a}})}\BibitemShut {NoStop}%
\bibitem [{\citenamefont {Yerokhin}\ \emph
  {et~al.}(2016{\natexlab{b}})\citenamefont {Yerokhin}, \citenamefont
  {Berseneva}, \citenamefont {Harman}, \citenamefont {Tupitsyn},\ and\
  \citenamefont {Keitel}}]{Yerokhin2016-1}%
  \BibitemOpen
  \bibfield  {author} {\bibinfo {author} {\bibfnamefont {V.~A.}\ \bibnamefont
  {Yerokhin}}, \bibinfo {author} {\bibfnamefont {E.}~\bibnamefont {Berseneva}},
  \bibinfo {author} {\bibfnamefont {Z.}~\bibnamefont {Harman}}, \bibinfo
  {author} {\bibfnamefont {I.~I.}\ \bibnamefont {Tupitsyn}}, \ and\ \bibinfo
  {author} {\bibfnamefont {C.~H.}\ \bibnamefont {Keitel}},\ }\href {\doibase
  10.1103/PhysRevA.94.022502} {\bibfield  {journal} {\bibinfo  {journal} {Phys.
  Rev. A}\ }\textbf {\bibinfo {volume} {94}},\ \bibinfo {pages} {022502}
  (\bibinfo {year} {2016}{\natexlab{b}})}\BibitemShut {NoStop}%
\bibitem [{\citenamefont {Volotka}\ and\ \citenamefont
  {Plunien}(2014)}]{Volotka2014-1}%
  \BibitemOpen
  \bibfield  {author} {\bibinfo {author} {\bibfnamefont {A.~V.}\ \bibnamefont
  {Volotka}}\ and\ \bibinfo {author} {\bibfnamefont {G.}~\bibnamefont
  {Plunien}},\ }\href {\doibase 10.1103/PhysRevLett.113.023002} {\bibfield
  {journal} {\bibinfo  {journal} {Phys. Rev. Lett.}\ }\textbf {\bibinfo
  {volume} {113}},\ \bibinfo {pages} {023002} (\bibinfo {year}
  {2014})}\BibitemShut {NoStop}%
\bibitem [{\citenamefont {Arapoglou}\ \emph {et~al.}(2019)\citenamefont
  {Arapoglou}, \citenamefont {Egl}, \citenamefont {H{\"o}cker}, \citenamefont
  {Sailer}, \citenamefont {Tu}, \citenamefont {Weigel}, \citenamefont {Wolf},
  \citenamefont {Cakir}, \citenamefont {Yerokhin}, \citenamefont {Oreshkina},
  \citenamefont {Agababaev}, \citenamefont {Volotka}, \citenamefont {Zinenko},
  \citenamefont {Glazov}, \citenamefont {Harman}, \citenamefont {Keitel},
  \citenamefont {Sturm},\ and\ \citenamefont {Blaum}}]{Arapoglou2019}%
  \BibitemOpen
  \bibfield  {author} {\bibinfo {author} {\bibfnamefont {I.}~\bibnamefont
  {Arapoglou}}, \bibinfo {author} {\bibfnamefont {A.}~\bibnamefont {Egl}},
  \bibinfo {author} {\bibfnamefont {M.}~\bibnamefont {H{\"o}cker}}, \bibinfo
  {author} {\bibfnamefont {T.}~\bibnamefont {Sailer}}, \bibinfo {author}
  {\bibfnamefont {B.}~\bibnamefont {Tu}}, \bibinfo {author} {\bibfnamefont
  {A.}~\bibnamefont {Weigel}}, \bibinfo {author} {\bibfnamefont
  {R.}~\bibnamefont {Wolf}}, \bibinfo {author} {\bibfnamefont {H.}~\bibnamefont
  {Cakir}}, \bibinfo {author} {\bibfnamefont {V.~A.}\ \bibnamefont {Yerokhin}},
  \bibinfo {author} {\bibfnamefont {N.~S.}\ \bibnamefont {Oreshkina}}, \bibinfo
  {author} {\bibfnamefont {V.~A.}\ \bibnamefont {Agababaev}}, \bibinfo {author}
  {\bibfnamefont {A.~V.}\ \bibnamefont {Volotka}}, \bibinfo {author}
  {\bibfnamefont {D.~V.}\ \bibnamefont {Zinenko}}, \bibinfo {author}
  {\bibfnamefont {D.~A.}\ \bibnamefont {Glazov}}, \bibinfo {author}
  {\bibfnamefont {Z.}~\bibnamefont {Harman}}, \bibinfo {author} {\bibfnamefont
  {C.~H.}\ \bibnamefont {Keitel}}, \bibinfo {author} {\bibfnamefont
  {S.}~\bibnamefont {Sturm}}, \ and\ \bibinfo {author} {\bibfnamefont
  {K.}~\bibnamefont {Blaum}},\ }\href {\doibase 10.1103/PhysRevLett.122.253001}
  {\bibfield  {journal} {\bibinfo  {journal} {Phys. Rev. Lett.}\ }\textbf
  {\bibinfo {volume} {122}},\ \bibinfo {pages} {253001} (\bibinfo {year}
  {2019})}\BibitemShut {NoStop}%
\bibitem [{\citenamefont {Shabaev}(2002)}]{Shabaev2002}%
  \BibitemOpen
  \bibfield  {author} {\bibinfo {author} {\bibfnamefont {V.}~\bibnamefont
  {Shabaev}},\ }\href {\doibase https://doi.org/10.1016/S0370-1573(01)00024-2}
  {\bibfield  {journal} {\bibinfo  {journal} {Phys. Rep.}\ }\textbf {\bibinfo
  {volume} {356}},\ \bibinfo {pages} {119 } (\bibinfo {year}
  {2002})}\BibitemShut {NoStop}%
\bibitem [{\citenamefont {Breit}(1928)}]{Breit1928}%
  \BibitemOpen
  \bibfield  {author} {\bibinfo {author} {\bibfnamefont {G.}~\bibnamefont
  {Breit}},\ }\href {https://doi.org/10.1038/122649a0} {\bibfield  {journal}
  {\bibinfo  {journal} {Nature}\ }\textbf {\bibinfo {volume} {122}},\ \bibinfo
  {pages} {649} (\bibinfo {year} {1928})}\BibitemShut {NoStop}%
\bibitem [{\citenamefont {Zapryagaev}(1979)}]{Zapryagaev1979}%
  \BibitemOpen
  \bibfield  {author} {\bibinfo {author} {\bibfnamefont {S.~A.}\ \bibnamefont
  {Zapryagaev}},\ }\href@noop {} {\bibfield  {journal} {\bibinfo  {journal}
  {Opt. Spektrosk.}\ }\textbf {\bibinfo {volume} {47}},\ \bibinfo {pages} {9}
  (\bibinfo {year} {1979})}\BibitemShut {NoStop}%
\bibitem [{\citenamefont {Angeli}\ and\ \citenamefont
  {Marinova}(2013)}]{Angeli2013}%
  \BibitemOpen
  \bibfield  {author} {\bibinfo {author} {\bibfnamefont {I.}~\bibnamefont
  {Angeli}}\ and\ \bibinfo {author} {\bibfnamefont {K.}~\bibnamefont
  {Marinova}},\ }\href {\doibase https://doi.org/10.1016/j.adt.2011.12.006}
  {\bibfield  {journal} {\bibinfo  {journal} {At. Data Nucl. Data Tables}\
  }\textbf {\bibinfo {volume} {99}},\ \bibinfo {pages} {69 } (\bibinfo {year}
  {2013})}\BibitemShut {NoStop}%
\bibitem [{\citenamefont {Shabaev}\ \emph {et~al.}(2002)\citenamefont
  {Shabaev}, \citenamefont {Glazov}, \citenamefont {Shabaeva}, \citenamefont
  {Yerokhin}, \citenamefont {Plunien},\ and\ \citenamefont
  {Soff}}]{Shabaev2002-1}%
  \BibitemOpen
  \bibfield  {author} {\bibinfo {author} {\bibfnamefont {V.~M.}\ \bibnamefont
  {Shabaev}}, \bibinfo {author} {\bibfnamefont {D.~A.}\ \bibnamefont {Glazov}},
  \bibinfo {author} {\bibfnamefont {M.~B.}\ \bibnamefont {Shabaeva}}, \bibinfo
  {author} {\bibfnamefont {V.~A.}\ \bibnamefont {Yerokhin}}, \bibinfo {author}
  {\bibfnamefont {G.}~\bibnamefont {Plunien}}, \ and\ \bibinfo {author}
  {\bibfnamefont {G.}~\bibnamefont {Soff}},\ }\href {\doibase
  10.1103/PhysRevA.65.062104} {\bibfield  {journal} {\bibinfo  {journal} {Phys.
  Rev. A}\ }\textbf {\bibinfo {volume} {65}},\ \bibinfo {pages} {062104}
  (\bibinfo {year} {2002})}\BibitemShut {NoStop}%
\bibitem [{\citenamefont {Johnson}\ \emph {et~al.}(1988)\citenamefont
  {Johnson}, \citenamefont {Blundell},\ and\ \citenamefont
  {Sapirstein}}]{Johnson1988}%
  \BibitemOpen
  \bibfield  {author} {\bibinfo {author} {\bibfnamefont {W.~R.}\ \bibnamefont
  {Johnson}}, \bibinfo {author} {\bibfnamefont {S.~A.}\ \bibnamefont
  {Blundell}}, \ and\ \bibinfo {author} {\bibfnamefont {J.}~\bibnamefont
  {Sapirstein}},\ }\href {\doibase 10.1103/PhysRevA.37.307} {\bibfield
  {journal} {\bibinfo  {journal} {Phys. Rev. A}\ }\textbf {\bibinfo {volume}
  {37}},\ \bibinfo {pages} {307} (\bibinfo {year} {1988})}\BibitemShut
  {NoStop}%
\bibitem [{\citenamefont {Shabaev}\ \emph {et~al.}(2004)\citenamefont
  {Shabaev}, \citenamefont {Tupitsyn}, \citenamefont {Yerokhin}, \citenamefont
  {Plunien},\ and\ \citenamefont {Soff}}]{Shabaev2004}%
  \BibitemOpen
  \bibfield  {author} {\bibinfo {author} {\bibfnamefont {V.~M.}\ \bibnamefont
  {Shabaev}}, \bibinfo {author} {\bibfnamefont {I.~I.}\ \bibnamefont
  {Tupitsyn}}, \bibinfo {author} {\bibfnamefont {V.~A.}\ \bibnamefont
  {Yerokhin}}, \bibinfo {author} {\bibfnamefont {G.}~\bibnamefont {Plunien}}, \
  and\ \bibinfo {author} {\bibfnamefont {G.}~\bibnamefont {Soff}},\ }\href
  {\doibase 10.1103/PhysRevLett.93.130405} {\bibfield  {journal} {\bibinfo
  {journal} {Phys. Rev. Lett.}\ }\textbf {\bibinfo {volume} {93}},\ \bibinfo
  {pages} {130405} (\bibinfo {year} {2004})}\BibitemShut {NoStop}%
\bibitem [{\citenamefont {Glazov}\ \emph {et~al.}(2013)\citenamefont {Glazov},
  \citenamefont {Volotka}, \citenamefont {Schepetnov}, \citenamefont {Sokolov},
  \citenamefont {Shabaev}, \citenamefont {Tupitsyn},\ and\ \citenamefont
  {Plunien}}]{Glazov2013}%
  \BibitemOpen
  \bibfield  {author} {\bibinfo {author} {\bibfnamefont {D.~A.}\ \bibnamefont
  {Glazov}}, \bibinfo {author} {\bibfnamefont {A.~V.}\ \bibnamefont {Volotka}},
  \bibinfo {author} {\bibfnamefont {A.~A.}\ \bibnamefont {Schepetnov}},
  \bibinfo {author} {\bibfnamefont {M.~M.}\ \bibnamefont {Sokolov}}, \bibinfo
  {author} {\bibfnamefont {V.~M.}\ \bibnamefont {Shabaev}}, \bibinfo {author}
  {\bibfnamefont {I.~I.}\ \bibnamefont {Tupitsyn}}, \ and\ \bibinfo {author}
  {\bibfnamefont {G.}~\bibnamefont {Plunien}},\ }\href {\doibase
  10.1088/0031-8949/2013/t156/014014} {\bibfield  {journal} {\bibinfo
  {journal} {Phys. Scr.}\ }\textbf {\bibinfo {volume} {T156}},\ \bibinfo
  {pages} {014014} (\bibinfo {year} {2013})}\BibitemShut {NoStop}%
\bibitem [{\citenamefont {Agababaev}\ \emph {et~al.}(2018)\citenamefont
  {Agababaev}, \citenamefont {Glazov}, \citenamefont {Volotka}, \citenamefont
  {Zinenko}, \citenamefont {Shabaev},\ and\ \citenamefont
  {Plunien}}]{Agababaev2018}%
  \BibitemOpen
  \bibfield  {author} {\bibinfo {author} {\bibfnamefont {V.~A.}\ \bibnamefont
  {Agababaev}}, \bibinfo {author} {\bibfnamefont {D.~A.}\ \bibnamefont
  {Glazov}}, \bibinfo {author} {\bibfnamefont {A.~V.}\ \bibnamefont {Volotka}},
  \bibinfo {author} {\bibfnamefont {D.~V.}\ \bibnamefont {Zinenko}}, \bibinfo
  {author} {\bibfnamefont {V.~M.}\ \bibnamefont {Shabaev}}, \ and\ \bibinfo
  {author} {\bibfnamefont {G.}~\bibnamefont {Plunien}},\ }\href {\doibase
  10.1088/1742-6596/1138/1/012003} {\bibfield  {journal} {\bibinfo  {journal}
  {J. Phys. Conf. Ser.}\ }\textbf {\bibinfo {volume} {1138}},\ \bibinfo {pages}
  {012003} (\bibinfo {year} {2018})}\BibitemShut {NoStop}%
\bibitem [{\citenamefont {Shchepetnov}\ \emph {et~al.}(2015)\citenamefont
  {Shchepetnov}, \citenamefont {Glazov}, \citenamefont {Volotka}, \citenamefont
  {Shabaev}, \citenamefont {Tupitsyn},\ and\ \citenamefont
  {Plunien}}]{Shchepetnov2015}%
  \BibitemOpen
  \bibfield  {author} {\bibinfo {author} {\bibfnamefont {A.~A.}\ \bibnamefont
  {Shchepetnov}}, \bibinfo {author} {\bibfnamefont {D.~A.}\ \bibnamefont
  {Glazov}}, \bibinfo {author} {\bibfnamefont {A.~V.}\ \bibnamefont {Volotka}},
  \bibinfo {author} {\bibfnamefont {V.~M.}\ \bibnamefont {Shabaev}}, \bibinfo
  {author} {\bibfnamefont {I.~I.}\ \bibnamefont {Tupitsyn}}, \ and\ \bibinfo
  {author} {\bibfnamefont {G.}~\bibnamefont {Plunien}},\ }\href {\doibase
  10.1088/1742-6596/583/1/012001} {\bibfield  {journal} {\bibinfo  {journal}
  {J. Phys. Conf. Ser.}\ }\textbf {\bibinfo {volume} {583}},\ \bibinfo {pages}
  {012001} (\bibinfo {year} {2015})}\BibitemShut {NoStop}%
\bibitem [{\citenamefont {Wagner}\ \emph {et~al.}(2013)\citenamefont {Wagner},
  \citenamefont {Sturm}, \citenamefont {K\"ohler}, \citenamefont {Glazov},
  \citenamefont {Volotka}, \citenamefont {Plunien}, \citenamefont {Quint},
  \citenamefont {Werth}, \citenamefont {Shabaev},\ and\ \citenamefont
  {Blaum}}]{Wagner2013}%
  \BibitemOpen
  \bibfield  {author} {\bibinfo {author} {\bibfnamefont {A.}~\bibnamefont
  {Wagner}}, \bibinfo {author} {\bibfnamefont {S.}~\bibnamefont {Sturm}},
  \bibinfo {author} {\bibfnamefont {F.}~\bibnamefont {K\"ohler}}, \bibinfo
  {author} {\bibfnamefont {D.~A.}\ \bibnamefont {Glazov}}, \bibinfo {author}
  {\bibfnamefont {A.~V.}\ \bibnamefont {Volotka}}, \bibinfo {author}
  {\bibfnamefont {G.}~\bibnamefont {Plunien}}, \bibinfo {author} {\bibfnamefont
  {W.}~\bibnamefont {Quint}}, \bibinfo {author} {\bibfnamefont
  {G.}~\bibnamefont {Werth}}, \bibinfo {author} {\bibfnamefont {V.~M.}\
  \bibnamefont {Shabaev}}, \ and\ \bibinfo {author} {\bibfnamefont
  {K.}~\bibnamefont {Blaum}},\ }\href {\doibase 10.1103/PhysRevLett.110.033003}
  {\bibfield  {journal} {\bibinfo  {journal} {Phys. Rev. Lett.}\ }\textbf
  {\bibinfo {volume} {110}},\ \bibinfo {pages} {033003} (\bibinfo {year}
  {2013})}\BibitemShut {NoStop}%
\bibitem [{\citenamefont {Yerokhin}\ and\ \citenamefont
  {Jentschura}(2010)}]{Yerokhin2010}%
  \BibitemOpen
  \bibfield  {author} {\bibinfo {author} {\bibfnamefont {V.~A.}\ \bibnamefont
  {Yerokhin}}\ and\ \bibinfo {author} {\bibfnamefont {U.~D.}\ \bibnamefont
  {Jentschura}},\ }\href {\doibase 10.1103/PhysRevA.81.012502} {\bibfield
  {journal} {\bibinfo  {journal} {Phys. Rev. A}\ }\textbf {\bibinfo {volume}
  {81}},\ \bibinfo {pages} {012502} (\bibinfo {year} {2010})}\BibitemShut
  {NoStop}%
\bibitem [{\citenamefont {Yerokhin}\ and\ \citenamefont
  {Harman}(2017)}]{Yerokhin2017}%
  \BibitemOpen
  \bibfield  {author} {\bibinfo {author} {\bibfnamefont {V.~A.}\ \bibnamefont
  {Yerokhin}}\ and\ \bibinfo {author} {\bibfnamefont {Z.}~\bibnamefont
  {Harman}},\ }\href {\doibase 10.1103/PhysRevA.95.060501} {\bibfield
  {journal} {\bibinfo  {journal} {Phys. Rev. A}\ }\textbf {\bibinfo {volume}
  {95}},\ \bibinfo {pages} {060501} (\bibinfo {year} {2017})}\BibitemShut
  {NoStop}%
\bibitem [{\citenamefont {Yerokhin}\ \emph {et~al.}(2013)\citenamefont
  {Yerokhin}, \citenamefont {Keitel},\ and\ \citenamefont
  {Harman}}]{Yerokhin2013}%
  \BibitemOpen
  \bibfield  {author} {\bibinfo {author} {\bibfnamefont {V.~A.}\ \bibnamefont
  {Yerokhin}}, \bibinfo {author} {\bibfnamefont {C.~H.}\ \bibnamefont
  {Keitel}}, \ and\ \bibinfo {author} {\bibfnamefont {Z.}~\bibnamefont
  {Harman}},\ }\href {\doibase 10.1088/0953-4075/46/24/245002} {\bibfield
  {journal} {\bibinfo  {journal} {J. Phys. B}\ }\textbf {\bibinfo {volume}
  {46}},\ \bibinfo {pages} {245002} (\bibinfo {year} {2013})}\BibitemShut
  {NoStop}%
\bibitem [{\citenamefont {Volotka}\ \emph {et~al.}(2009)\citenamefont
  {Volotka}, \citenamefont {Glazov}, \citenamefont {Shabaev}, \citenamefont
  {Tupitsyn},\ and\ \citenamefont {Plunien}}]{Volotka2009}%
  \BibitemOpen
  \bibfield  {author} {\bibinfo {author} {\bibfnamefont {A.~V.}\ \bibnamefont
  {Volotka}}, \bibinfo {author} {\bibfnamefont {D.~A.}\ \bibnamefont {Glazov}},
  \bibinfo {author} {\bibfnamefont {V.~M.}\ \bibnamefont {Shabaev}}, \bibinfo
  {author} {\bibfnamefont {I.~I.}\ \bibnamefont {Tupitsyn}}, \ and\ \bibinfo
  {author} {\bibfnamefont {G.}~\bibnamefont {Plunien}},\ }\href {\doibase
  10.1103/PhysRevLett.103.033005} {\bibfield  {journal} {\bibinfo  {journal}
  {Phys. Rev. Lett.}\ }\textbf {\bibinfo {volume} {103}},\ \bibinfo {pages}
  {033005} (\bibinfo {year} {2009})}\BibitemShut {NoStop}%
\bibitem [{\citenamefont {Volotka}\ \emph {et~al.}(2014)\citenamefont
  {Volotka}, \citenamefont {Glazov}, \citenamefont {Shabaev}, \citenamefont
  {Tupitsyn},\ and\ \citenamefont {Plunien}}]{Volotka2014}%
  \BibitemOpen
  \bibfield  {author} {\bibinfo {author} {\bibfnamefont {A.~V.}\ \bibnamefont
  {Volotka}}, \bibinfo {author} {\bibfnamefont {D.~A.}\ \bibnamefont {Glazov}},
  \bibinfo {author} {\bibfnamefont {V.~M.}\ \bibnamefont {Shabaev}}, \bibinfo
  {author} {\bibfnamefont {I.~I.}\ \bibnamefont {Tupitsyn}}, \ and\ \bibinfo
  {author} {\bibfnamefont {G.}~\bibnamefont {Plunien}},\ }\href {\doibase
  10.1103/PhysRevLett.112.253004} {\bibfield  {journal} {\bibinfo  {journal}
  {Phys. Rev. Lett.}\ }\textbf {\bibinfo {volume} {112}},\ \bibinfo {pages}
  {253004} (\bibinfo {year} {2014})}\BibitemShut {NoStop}%
\bibitem [{\citenamefont {Glazov}\ \emph {et~al.}(2006)\citenamefont {Glazov},
  \citenamefont {Volotka}, \citenamefont {Shabaev}, \citenamefont {Tupitsyn},\
  and\ \citenamefont {Plunien}}]{Glazov2006}%
  \BibitemOpen
  \bibfield  {author} {\bibinfo {author} {\bibfnamefont {D.~A.}\ \bibnamefont
  {Glazov}}, \bibinfo {author} {\bibfnamefont {A.~V.}\ \bibnamefont {Volotka}},
  \bibinfo {author} {\bibfnamefont {V.~M.}\ \bibnamefont {Shabaev}}, \bibinfo
  {author} {\bibfnamefont {I.~I.}\ \bibnamefont {Tupitsyn}}, \ and\ \bibinfo
  {author} {\bibfnamefont {G.}~\bibnamefont {Plunien}},\ }\href {\doibase
  https://doi.org/10.1016/j.physleta.2006.04.056} {\bibfield  {journal}
  {\bibinfo  {journal} {Phys. Lett. A}\ }\textbf {\bibinfo {volume} {357}},\
  \bibinfo {pages} {330 } (\bibinfo {year} {2006})}\BibitemShut {NoStop}%
\bibitem [{\citenamefont {Glazov}\ \emph {et~al.}(2004)\citenamefont {Glazov},
  \citenamefont {Shabaev}, \citenamefont {Tupitsyn}, \citenamefont {Volotka},
  \citenamefont {Yerokhin}, \citenamefont {Plunien},\ and\ \citenamefont
  {Soff}}]{Glazov2004}%
  \BibitemOpen
  \bibfield  {author} {\bibinfo {author} {\bibfnamefont {D.~A.}\ \bibnamefont
  {Glazov}}, \bibinfo {author} {\bibfnamefont {V.~M.}\ \bibnamefont {Shabaev}},
  \bibinfo {author} {\bibfnamefont {I.~I.}\ \bibnamefont {Tupitsyn}}, \bibinfo
  {author} {\bibfnamefont {A.~V.}\ \bibnamefont {Volotka}}, \bibinfo {author}
  {\bibfnamefont {V.~A.}\ \bibnamefont {Yerokhin}}, \bibinfo {author}
  {\bibfnamefont {G.}~\bibnamefont {Plunien}}, \ and\ \bibinfo {author}
  {\bibfnamefont {G.}~\bibnamefont {Soff}},\ }\href {\doibase
  10.1103/PhysRevA.70.062104} {\bibfield  {journal} {\bibinfo  {journal} {Phys.
  Rev. A}\ }\textbf {\bibinfo {volume} {70}},\ \bibinfo {pages} {062104}
  (\bibinfo {year} {2004})}\BibitemShut {NoStop}%
\bibitem [{\citenamefont {Yerokhin}\ \emph {et~al.}(2007)\citenamefont
  {Yerokhin}, \citenamefont {Artemyev},\ and\ \citenamefont
  {Shabaev}}]{Yerokhin2007}%
  \BibitemOpen
  \bibfield  {author} {\bibinfo {author} {\bibfnamefont {V.~A.}\ \bibnamefont
  {Yerokhin}}, \bibinfo {author} {\bibfnamefont {A.~N.}\ \bibnamefont
  {Artemyev}}, \ and\ \bibinfo {author} {\bibfnamefont {V.~M.}\ \bibnamefont
  {Shabaev}},\ }\href {\doibase 10.1103/PhysRevA.75.062501} {\bibfield
  {journal} {\bibinfo  {journal} {Phys. Rev. A}\ }\textbf {\bibinfo {volume}
  {75}},\ \bibinfo {pages} {062501} (\bibinfo {year} {2007})}\BibitemShut
  {NoStop}%
\bibitem [{\citenamefont {Yerokhin}(2011)}]{Yerokhin2011}%
  \BibitemOpen
  \bibfield  {author} {\bibinfo {author} {\bibfnamefont {V.~A.}\ \bibnamefont
  {Yerokhin}},\ }\href {\doibase 10.1103/PhysRevA.83.012507} {\bibfield
  {journal} {\bibinfo  {journal} {Phys. Rev. A}\ }\textbf {\bibinfo {volume}
  {83}},\ \bibinfo {pages} {012507} (\bibinfo {year} {2011})}\BibitemShut
  {NoStop}%
\bibitem [{\citenamefont {Hegstrom}(1973)}]{Hegstrom1973}%
  \BibitemOpen
  \bibfield  {author} {\bibinfo {author} {\bibfnamefont {R.~A.}\ \bibnamefont
  {Hegstrom}},\ }\href {\doibase 10.1103/PhysRevA.7.451} {\bibfield  {journal}
  {\bibinfo  {journal} {Phys. Rev. A}\ }\textbf {\bibinfo {volume} {7}},\
  \bibinfo {pages} {451} (\bibinfo {year} {1973})}\BibitemShut {NoStop}%
\bibitem [{\citenamefont {Soff}\ and\ \citenamefont {Mohr}(1988)}]{Soff1988}%
  \BibitemOpen
  \bibfield  {author} {\bibinfo {author} {\bibfnamefont {G.}~\bibnamefont
  {Soff}}\ and\ \bibinfo {author} {\bibfnamefont {P.~J.}\ \bibnamefont
  {Mohr}},\ }\href {\doibase 10.1103/PhysRevA.38.5066} {\bibfield  {journal}
  {\bibinfo  {journal} {Phys. Rev. A}\ }\textbf {\bibinfo {volume} {38}},\
  \bibinfo {pages} {5066} (\bibinfo {year} {1988})}\BibitemShut {NoStop}%
\bibitem [{\citenamefont {Fullerton}\ and\ \citenamefont
  {Rinker}(1976)}]{Fullerton1976}%
  \BibitemOpen
  \bibfield  {author} {\bibinfo {author} {\bibfnamefont {L.~W.}\ \bibnamefont
  {Fullerton}}\ and\ \bibinfo {author} {\bibfnamefont {G.~A.}\ \bibnamefont
  {Rinker}},\ }\href {\doibase 10.1103/PhysRevA.13.1283} {\bibfield  {journal}
  {\bibinfo  {journal} {Phys. Rev. A}\ }\textbf {\bibinfo {volume} {13}},\
  \bibinfo {pages} {1283} (\bibinfo {year} {1976})}\BibitemShut {NoStop}%
\bibitem [{\citenamefont {Klarsfeld}(1977)}]{Klarsfeld1977}%
  \BibitemOpen
  \bibfield  {author} {\bibinfo {author} {\bibfnamefont {S.}~\bibnamefont
  {Klarsfeld}},\ }\href {\doibase https://doi.org/10.1016/0370-2693(77)90620-7}
  {\bibfield  {journal} {\bibinfo  {journal} {Phys. Lett. B}\ }\textbf
  {\bibinfo {volume} {66}},\ \bibinfo {pages} {86 } (\bibinfo {year}
  {1977})}\BibitemShut {NoStop}%
\bibitem [{\citenamefont {Wichmann}\ and\ \citenamefont
  {Kroll}(1956)}]{Wichmann1956}%
  \BibitemOpen
  \bibfield  {author} {\bibinfo {author} {\bibfnamefont {E.~H.}\ \bibnamefont
  {Wichmann}}\ and\ \bibinfo {author} {\bibfnamefont {N.~M.}\ \bibnamefont
  {Kroll}},\ }\href {\doibase 10.1103/PhysRev.101.843} {\bibfield  {journal}
  {\bibinfo  {journal} {Phys. Rev.}\ }\textbf {\bibinfo {volume} {101}},\
  \bibinfo {pages} {843} (\bibinfo {year} {1956})}\BibitemShut {NoStop}%
\bibitem [{\citenamefont {Salvat}\ and\ \citenamefont
  {Mayol}(1991)}]{Salvat1991}%
  \BibitemOpen
  \bibfield  {author} {\bibinfo {author} {\bibfnamefont {F.}~\bibnamefont
  {Salvat}}\ and\ \bibinfo {author} {\bibfnamefont {R.}~\bibnamefont {Mayol}},\
  }\href {\doibase https://doi.org/10.1016/0010-4655(91)90122-2} {\bibfield
  {journal} {\bibinfo  {journal} {Comput. Phys. Commun.}\ }\textbf {\bibinfo
  {volume} {62}},\ \bibinfo {pages} {65 } (\bibinfo {year} {1991})}\BibitemShut
  {NoStop}%
\bibitem [{\citenamefont {Salvat}\ \emph {et~al.}(1995)\citenamefont {Salvat},
  \citenamefont {Fernández-Varea},\ and\ \citenamefont
  {Williamson}}]{Salvat1995}%
  \BibitemOpen
  \bibfield  {author} {\bibinfo {author} {\bibfnamefont {F.}~\bibnamefont
  {Salvat}}, \bibinfo {author} {\bibfnamefont {J.}~\bibnamefont
  {Fernández-Varea}}, \ and\ \bibinfo {author} {\bibfnamefont
  {W.}~\bibnamefont {Williamson}},\ }\href {\doibase
  https://doi.org/10.1016/0010-4655(95)00039-I} {\bibfield  {journal} {\bibinfo
   {journal} {Comput. Phys. Commun.}\ }\textbf {\bibinfo {volume} {90}},\
  \bibinfo {pages} {151 } (\bibinfo {year} {1995})}\BibitemShut {NoStop}%
\bibitem [{\citenamefont {Fainshtein}\ \emph {et~al.}(1991)\citenamefont
  {Fainshtein}, \citenamefont {Manakov},\ and\ \citenamefont
  {Nekipelov}}]{Fainshtein1991}%
  \BibitemOpen
  \bibfield  {author} {\bibinfo {author} {\bibfnamefont {A.~G.}\ \bibnamefont
  {Fainshtein}}, \bibinfo {author} {\bibfnamefont {N.~L.}\ \bibnamefont
  {Manakov}}, \ and\ \bibinfo {author} {\bibfnamefont {A.~A.}\ \bibnamefont
  {Nekipelov}},\ }\href {\doibase 10.1088/0953-4075/24/3/012} {\bibfield
  {journal} {\bibinfo  {journal} {J. Phys. B}\ }\textbf {\bibinfo {volume}
  {24}},\ \bibinfo {pages} {559} (\bibinfo {year} {1991})}\BibitemShut
  {NoStop}%
\bibitem [{\citenamefont {Lee}\ \emph {et~al.}(2007)\citenamefont {Lee},
  \citenamefont {Milstein}, \citenamefont {Terekhov},\ and\ \citenamefont
  {Karshenboim}}]{Lee2007}%
  \BibitemOpen
  \bibfield  {author} {\bibinfo {author} {\bibfnamefont {R.~N.}\ \bibnamefont
  {Lee}}, \bibinfo {author} {\bibfnamefont {A.~I.}\ \bibnamefont {Milstein}},
  \bibinfo {author} {\bibfnamefont {I.~S.}\ \bibnamefont {Terekhov}}, \ and\
  \bibinfo {author} {\bibfnamefont {S.~G.}\ \bibnamefont {Karshenboim}},\
  }\href {\doibase 10.1139/p07-024} {\bibfield  {journal} {\bibinfo  {journal}
  {Can. J. Phys.}\ }\textbf {\bibinfo {volume} {85}},\ \bibinfo {pages} {541}
  (\bibinfo {year} {2007})},\ \Eprint
  {http://arxiv.org/abs/https://doi.org/10.1139/p07-024}
  {https://doi.org/10.1139/p07-024} \BibitemShut {NoStop}%
\bibitem [{\citenamefont {Glazov}\ \emph {et~al.}(2010)\citenamefont {Glazov},
  \citenamefont {Volotka}, \citenamefont {Shabaev}, \citenamefont {Tupitsyn},\
  and\ \citenamefont {Plunien}}]{Glazov2010}%
  \BibitemOpen
  \bibfield  {author} {\bibinfo {author} {\bibfnamefont {D.~A.}\ \bibnamefont
  {Glazov}}, \bibinfo {author} {\bibfnamefont {A.~V.}\ \bibnamefont {Volotka}},
  \bibinfo {author} {\bibfnamefont {V.~M.}\ \bibnamefont {Shabaev}}, \bibinfo
  {author} {\bibfnamefont {I.~I.}\ \bibnamefont {Tupitsyn}}, \ and\ \bibinfo
  {author} {\bibfnamefont {G.}~\bibnamefont {Plunien}},\ }\href {\doibase
  10.1103/PhysRevA.81.062112} {\bibfield  {journal} {\bibinfo  {journal} {Phys.
  Rev. A}\ }\textbf {\bibinfo {volume} {81}},\ \bibinfo {pages} {062112}
  (\bibinfo {year} {2010})}\BibitemShut {NoStop}%
\bibitem [{\citenamefont {Artemyev}\ \emph {et~al.}(1997)\citenamefont
  {Artemyev}, \citenamefont {Shabaev},\ and\ \citenamefont
  {Yerokhin}}]{Artemyev1997}%
  \BibitemOpen
  \bibfield  {author} {\bibinfo {author} {\bibfnamefont {A.~N.}\ \bibnamefont
  {Artemyev}}, \bibinfo {author} {\bibfnamefont {V.~M.}\ \bibnamefont
  {Shabaev}}, \ and\ \bibinfo {author} {\bibfnamefont {V.~A.}\ \bibnamefont
  {Yerokhin}},\ }\href {\doibase 10.1103/PhysRevA.56.3529} {\bibfield
  {journal} {\bibinfo  {journal} {Phys. Rev. A}\ }\textbf {\bibinfo {volume}
  {56}},\ \bibinfo {pages} {3529} (\bibinfo {year} {1997})}\BibitemShut
  {NoStop}%
\bibitem [{\citenamefont {Artemyev}\ \emph {et~al.}(1999)\citenamefont
  {Artemyev}, \citenamefont {Beier}, \citenamefont {Plunien}, \citenamefont
  {Shabaev}, \citenamefont {Soff},\ and\ \citenamefont
  {Yerokhin}}]{Artemyev1999}%
  \BibitemOpen
  \bibfield  {author} {\bibinfo {author} {\bibfnamefont {A.~N.}\ \bibnamefont
  {Artemyev}}, \bibinfo {author} {\bibfnamefont {T.}~\bibnamefont {Beier}},
  \bibinfo {author} {\bibfnamefont {G.}~\bibnamefont {Plunien}}, \bibinfo
  {author} {\bibfnamefont {V.~M.}\ \bibnamefont {Shabaev}}, \bibinfo {author}
  {\bibfnamefont {G.}~\bibnamefont {Soff}}, \ and\ \bibinfo {author}
  {\bibfnamefont {V.~A.}\ \bibnamefont {Yerokhin}},\ }\href {\doibase
  10.1103/PhysRevA.60.45} {\bibfield  {journal} {\bibinfo  {journal} {Phys.
  Rev. A}\ }\textbf {\bibinfo {volume} {60}},\ \bibinfo {pages} {45} (\bibinfo
  {year} {1999})}\BibitemShut {NoStop}%
\bibitem [{\citenamefont {Karshenboim}\ \emph {et~al.}(2001)\citenamefont
  {Karshenboim}, \citenamefont {Ivanov},\ and\ \citenamefont
  {Shabaev}}]{Karshenboim2001}%
  \BibitemOpen
  \bibfield  {author} {\bibinfo {author} {\bibfnamefont {S.~G.}\ \bibnamefont
  {Karshenboim}}, \bibinfo {author} {\bibfnamefont {V.~G.}\ \bibnamefont
  {Ivanov}}, \ and\ \bibinfo {author} {\bibfnamefont {V.~M.}\ \bibnamefont
  {Shabaev}},\ }\href {\doibase 10.1134/1.1410592} {\bibfield  {journal}
  {\bibinfo  {journal} {J. Exp. Theor. Phys.}\ }\textbf {\bibinfo {volume}
  {93}},\ \bibinfo {pages} {477} (\bibinfo {year} {2001})}\BibitemShut
  {NoStop}%
\bibitem [{\citenamefont {Shabaev}(2003)}]{Shabaev2003}%
  \BibitemOpen
  \bibfield  {author} {\bibinfo {author} {\bibfnamefont {V.~M.}\ \bibnamefont
  {Shabaev}},\ }\enquote {\bibinfo {title} {{Virial Relations for the Dirac
  Equation and Their Applications to Calculations of Hydrogen-Like Atoms}},}\
  in\ \href {\doibase 10.1007/978-3-540-45059-7_6} {\emph {\bibinfo {booktitle}
  {Precision Physics of Simple Atomic Systems}}},\ \bibinfo {editor} {edited
  by\ \bibinfo {editor} {\bibfnamefont {S.~G.}\ \bibnamefont {Karshenboim}}\
  and\ \bibinfo {editor} {\bibfnamefont {V.~B.}\ \bibnamefont {Smirnov}}}\
  (\bibinfo  {publisher} {Springer Berlin Heidelberg},\ \bibinfo {address}
  {Berlin, Heidelberg},\ \bibinfo {year} {2003})\ pp.\ \bibinfo {pages}
  {97--113}\BibitemShut {NoStop}%
\bibitem [{\citenamefont {Peskin}\ and\ \citenamefont
  {Schroeder}(1995)}]{Peskin1995}%
  \BibitemOpen
  \bibfield  {author} {\bibinfo {author} {\bibfnamefont {M.~E.}\ \bibnamefont
  {Peskin}}\ and\ \bibinfo {author} {\bibfnamefont {D.~V.}\ \bibnamefont
  {Schroeder}},\ }\href@noop {} {\emph {\bibinfo {title} {An Introduction to
  Quantum Field Theory}}}\ (\bibinfo  {publisher} {Westview Press},\ \bibinfo
  {year} {1995})\BibitemShut {NoStop}%
\bibitem [{\citenamefont {Shabaev}\ \emph {et~al.}(2018)\citenamefont
  {Shabaev}, \citenamefont {Glazov}, \citenamefont {Malyshev},\ and\
  \citenamefont {Tupitsyn}}]{Shabaev2018}%
  \BibitemOpen
  \bibfield  {author} {\bibinfo {author} {\bibfnamefont {V.~M.}\ \bibnamefont
  {Shabaev}}, \bibinfo {author} {\bibfnamefont {D.~A.}\ \bibnamefont {Glazov}},
  \bibinfo {author} {\bibfnamefont {A.~V.}\ \bibnamefont {Malyshev}}, \ and\
  \bibinfo {author} {\bibfnamefont {I.~I.}\ \bibnamefont {Tupitsyn}},\ }\href
  {\doibase 10.1103/PhysRevA.98.032512} {\bibfield  {journal} {\bibinfo
  {journal} {Phys. Rev. A}\ }\textbf {\bibinfo {volume} {98}},\ \bibinfo
  {pages} {032512} (\bibinfo {year} {2018})}\BibitemShut {NoStop}%
\bibitem [{\citenamefont {Glazov}\ \emph {et~al.}(2018)\citenamefont {Glazov},
  \citenamefont {Malyshev}, \citenamefont {Shabaev},\ and\ \citenamefont
  {Tupitsyn}}]{Glazov2018}%
  \BibitemOpen
  \bibfield  {author} {\bibinfo {author} {\bibfnamefont {D.~A.}\ \bibnamefont
  {Glazov}}, \bibinfo {author} {\bibfnamefont {A.~V.}\ \bibnamefont
  {Malyshev}}, \bibinfo {author} {\bibfnamefont {V.~M.}\ \bibnamefont
  {Shabaev}}, \ and\ \bibinfo {author} {\bibfnamefont {I.~I.}\ \bibnamefont
  {Tupitsyn}},\ }\href {\doibase 10.1134/S0030400X18040082} {\bibfield
  {journal} {\bibinfo  {journal} {Opt. Spectrosc.}\ }\textbf {\bibinfo {volume}
  {124}},\ \bibinfo {pages} {457} (\bibinfo {year} {2018})}\BibitemShut
  {NoStop}%
\bibitem [{\citenamefont {Aleksandrov}\ \emph {et~al.}(2018)\citenamefont
  {Aleksandrov}, \citenamefont {Glazov}, \citenamefont {Malyshev},
  \citenamefont {Shabaev},\ and\ \citenamefont {Tupitsyn}}]{Aleksandrov2018}%
  \BibitemOpen
  \bibfield  {author} {\bibinfo {author} {\bibfnamefont {I.~A.}\ \bibnamefont
  {Aleksandrov}}, \bibinfo {author} {\bibfnamefont {D.~A.}\ \bibnamefont
  {Glazov}}, \bibinfo {author} {\bibfnamefont {A.~V.}\ \bibnamefont
  {Malyshev}}, \bibinfo {author} {\bibfnamefont {V.~M.}\ \bibnamefont
  {Shabaev}}, \ and\ \bibinfo {author} {\bibfnamefont {I.~I.}\ \bibnamefont
  {Tupitsyn}},\ }\href {\doibase 10.1103/PhysRevA.98.062521} {\bibfield
  {journal} {\bibinfo  {journal} {Phys. Rev. A}\ }\textbf {\bibinfo {volume}
  {98}},\ \bibinfo {pages} {062521} (\bibinfo {year} {2018})}\BibitemShut
  {NoStop}%
\bibitem [{\citenamefont {Glazov}\ \emph {et~al.}(2020)\citenamefont {Glazov},
  \citenamefont {Malyshev}, \citenamefont {Shabaev},\ and\ \citenamefont
  {Tupitsyn}}]{Glazov2020}%
  \BibitemOpen
  \bibfield  {author} {\bibinfo {author} {\bibfnamefont {D.~A.}\ \bibnamefont
  {Glazov}}, \bibinfo {author} {\bibfnamefont {A.~V.}\ \bibnamefont
  {Malyshev}}, \bibinfo {author} {\bibfnamefont {V.~M.}\ \bibnamefont
  {Shabaev}}, \ and\ \bibinfo {author} {\bibfnamefont {I.~I.}\ \bibnamefont
  {Tupitsyn}},\ }\href {\doibase 10.1103/PhysRevA.101.012515} {\bibfield
  {journal} {\bibinfo  {journal} {Phys. Rev. A}\ }\textbf {\bibinfo {volume}
  {101}},\ \bibinfo {pages} {012515} (\bibinfo {year} {2020})}\BibitemShut
  {NoStop}%
\bibitem [{\citenamefont {Grotch}\ and\ \citenamefont
  {Kashuba}(1973)}]{Grotch1973}%
  \BibitemOpen
  \bibfield  {author} {\bibinfo {author} {\bibfnamefont {H.}~\bibnamefont
  {Grotch}}\ and\ \bibinfo {author} {\bibfnamefont {R.}~\bibnamefont
  {Kashuba}},\ }\href {\doibase 10.1103/PhysRevA.7.78} {\bibfield  {journal}
  {\bibinfo  {journal} {Phys. Rev. A}\ }\textbf {\bibinfo {volume} {7}},\
  \bibinfo {pages} {78} (\bibinfo {year} {1973})}\BibitemShut {NoStop}%
\bibitem [{\citenamefont {Mohr}\ \emph {et~al.}(2016)\citenamefont {Mohr},
  \citenamefont {Newell},\ and\ \citenamefont {Taylor}}]{Mohr2016}%
  \BibitemOpen
  \bibfield  {author} {\bibinfo {author} {\bibfnamefont {P.~J.}\ \bibnamefont
  {Mohr}}, \bibinfo {author} {\bibfnamefont {D.~B.}\ \bibnamefont {Newell}}, \
  and\ \bibinfo {author} {\bibfnamefont {B.~N.}\ \bibnamefont {Taylor}},\
  }\href {\doibase 10.1103/RevModPhys.88.035009} {\bibfield  {journal}
  {\bibinfo  {journal} {Rev. Mod. Phys.}\ }\textbf {\bibinfo {volume} {88}},\
  \bibinfo {pages} {035009} (\bibinfo {year} {2016})}\BibitemShut {NoStop}%
\bibitem [{\citenamefont {Marques}\ \emph {et~al.}(2016)\citenamefont
  {Marques}, \citenamefont {Indelicato}, \citenamefont {Parente}, \citenamefont
  {Sampaio},\ and\ \citenamefont {Santos}}]{Marques2016}%
  \BibitemOpen
  \bibfield  {author} {\bibinfo {author} {\bibfnamefont {J.~P.}\ \bibnamefont
  {Marques}}, \bibinfo {author} {\bibfnamefont {P.}~\bibnamefont {Indelicato}},
  \bibinfo {author} {\bibfnamefont {F.}~\bibnamefont {Parente}}, \bibinfo
  {author} {\bibfnamefont {J.~M.}\ \bibnamefont {Sampaio}}, \ and\ \bibinfo
  {author} {\bibfnamefont {J.~P.}\ \bibnamefont {Santos}},\ }\href {\doibase
  10.1103/PhysRevA.94.042504} {\bibfield  {journal} {\bibinfo  {journal} {Phys.
  Rev. A}\ }\textbf {\bibinfo {volume} {94}},\ \bibinfo {pages} {042504}
  (\bibinfo {year} {2016})}\BibitemShut {NoStop}%
\bibitem [{\citenamefont {Verdebout}\ \emph {et~al.}(2014)\citenamefont
  {Verdebout}, \citenamefont {Naz\'e}, \citenamefont {J{\"o}nsson},
  \citenamefont {Rynkun}, \citenamefont {Godefroid},\ and\ \citenamefont
  {Gaigalas}}]{Verdebout2014}%
  \BibitemOpen
  \bibfield  {author} {\bibinfo {author} {\bibfnamefont {S.}~\bibnamefont
  {Verdebout}}, \bibinfo {author} {\bibfnamefont {C.}~\bibnamefont {Naz\'e}},
  \bibinfo {author} {\bibfnamefont {P.}~\bibnamefont {J{\"o}nsson}}, \bibinfo
  {author} {\bibfnamefont {P.}~\bibnamefont {Rynkun}}, \bibinfo {author}
  {\bibfnamefont {M.}~\bibnamefont {Godefroid}}, \ and\ \bibinfo {author}
  {\bibfnamefont {G.}~\bibnamefont {Gaigalas}},\ }\href@noop {} {\bibfield
  {journal} {\bibinfo  {journal} {At. Data Nucl. Data Tables}\ }\textbf
  {\bibinfo {volume} {100}},\ \bibinfo {pages} {1111 } (\bibinfo {year}
  {2014})}\BibitemShut {NoStop}%
\bibitem [{\citenamefont {K\"{o}hler}\ \emph {et~al.}(2016)\citenamefont
  {K\"{o}hler}, \citenamefont {Blaum}, \citenamefont {Block}, \citenamefont
  {Chenmarev}, \citenamefont {Eliseev}, \citenamefont {Glazov}, \citenamefont
  {Goncharov}, \citenamefont {Hou}, \citenamefont {Kracke}, \citenamefont
  {Nesterenko}, \citenamefont {Novikov}, \citenamefont {Quint}, \citenamefont
  {Minaya~Ramirez}, \citenamefont {Shabaev}, \citenamefont {Sturm},
  \citenamefont {Volotka},\ and\ \citenamefont {Werth}}]{Koehler2016}%
  \BibitemOpen
  \bibfield  {author} {\bibinfo {author} {\bibfnamefont {F.}~\bibnamefont
  {K\"{o}hler}}, \bibinfo {author} {\bibfnamefont {K.}~\bibnamefont {Blaum}},
  \bibinfo {author} {\bibfnamefont {M.}~\bibnamefont {Block}}, \bibinfo
  {author} {\bibfnamefont {S.}~\bibnamefont {Chenmarev}}, \bibinfo {author}
  {\bibfnamefont {S.}~\bibnamefont {Eliseev}}, \bibinfo {author} {\bibfnamefont
  {D.~A.}\ \bibnamefont {Glazov}}, \bibinfo {author} {\bibfnamefont
  {M.}~\bibnamefont {Goncharov}}, \bibinfo {author} {\bibfnamefont
  {J.}~\bibnamefont {Hou}}, \bibinfo {author} {\bibfnamefont {A.}~\bibnamefont
  {Kracke}}, \bibinfo {author} {\bibfnamefont {D.~A.}\ \bibnamefont
  {Nesterenko}}, \bibinfo {author} {\bibfnamefont {Y.~N.}\ \bibnamefont
  {Novikov}}, \bibinfo {author} {\bibfnamefont {W.}~\bibnamefont {Quint}},
  \bibinfo {author} {\bibfnamefont {E.}~\bibnamefont {Minaya~Ramirez}},
  \bibinfo {author} {\bibfnamefont {V.~M.}\ \bibnamefont {Shabaev}}, \bibinfo
  {author} {\bibfnamefont {S.}~\bibnamefont {Sturm}}, \bibinfo {author}
  {\bibfnamefont {A.~V.}\ \bibnamefont {Volotka}}, \ and\ \bibinfo {author}
  {\bibfnamefont {G.}~\bibnamefont {Werth}},\ }\href {\doibase
  10.1038/ncomms10246} {\bibfield  {journal} {\bibinfo  {journal} {Nat.
  Commun.}\ }\textbf {\bibinfo {volume} {7}},\ \bibinfo {pages} {10246}
  (\bibinfo {year} {2016})}\BibitemShut {NoStop}%
\bibitem [{\citenamefont {Yerokhin}\ \emph {et~al.}(2017)\citenamefont
  {Yerokhin}, \citenamefont {Pachucki}, \citenamefont {Puchalski},
  \citenamefont {Harman},\ and\ \citenamefont {Keitel}}]{Yerokhin2017-1}%
  \BibitemOpen
  \bibfield  {author} {\bibinfo {author} {\bibfnamefont {V.~A.}\ \bibnamefont
  {Yerokhin}}, \bibinfo {author} {\bibfnamefont {K.}~\bibnamefont {Pachucki}},
  \bibinfo {author} {\bibfnamefont {M.}~\bibnamefont {Puchalski}}, \bibinfo
  {author} {\bibfnamefont {Z.}~\bibnamefont {Harman}}, \ and\ \bibinfo {author}
  {\bibfnamefont {C.~H.}\ \bibnamefont {Keitel}},\ }\href {\doibase
  10.1103/PhysRevA.95.062511} {\bibfield  {journal} {\bibinfo  {journal} {Phys.
  Rev. A}\ }\textbf {\bibinfo {volume} {95}},\ \bibinfo {pages} {062511}
  (\bibinfo {year} {2017})}\BibitemShut {NoStop}%
\bibitem [{\citenamefont {Glazov}\ \emph {et~al.}(2019)\citenamefont {Glazov},
  \citenamefont {K\"ohler-Langes}, \citenamefont {Volotka}, \citenamefont
  {Blaum}, \citenamefont {Hei\ss{}e}, \citenamefont {Plunien}, \citenamefont
  {Quint}, \citenamefont {Rau}, \citenamefont {Shabaev}, \citenamefont
  {Sturm},\ and\ \citenamefont {Werth}}]{Glazov2019}%
  \BibitemOpen
  \bibfield  {author} {\bibinfo {author} {\bibfnamefont {D.~A.}\ \bibnamefont
  {Glazov}}, \bibinfo {author} {\bibfnamefont {F.}~\bibnamefont
  {K\"ohler-Langes}}, \bibinfo {author} {\bibfnamefont {A.~V.}\ \bibnamefont
  {Volotka}}, \bibinfo {author} {\bibfnamefont {K.}~\bibnamefont {Blaum}},
  \bibinfo {author} {\bibfnamefont {F.}~\bibnamefont {Hei\ss{}e}}, \bibinfo
  {author} {\bibfnamefont {G.}~\bibnamefont {Plunien}}, \bibinfo {author}
  {\bibfnamefont {W.}~\bibnamefont {Quint}}, \bibinfo {author} {\bibfnamefont
  {S.}~\bibnamefont {Rau}}, \bibinfo {author} {\bibfnamefont {V.~M.}\
  \bibnamefont {Shabaev}}, \bibinfo {author} {\bibfnamefont {S.}~\bibnamefont
  {Sturm}}, \ and\ \bibinfo {author} {\bibfnamefont {G.}~\bibnamefont
  {Werth}},\ }\href {\doibase 10.1103/PhysRevLett.123.173001} {\bibfield
  {journal} {\bibinfo  {journal} {Phys. Rev. Lett.}\ }\textbf {\bibinfo
  {volume} {123}},\ \bibinfo {pages} {173001} (\bibinfo {year}
  {2019})}\BibitemShut {NoStop}%
\bibitem [{\citenamefont {Sturm}\ \emph {et~al.}(2019)\citenamefont {Sturm},
  \citenamefont {Arapoglou}, \citenamefont {Egl}, \citenamefont {H\"{o}cker},
  \citenamefont {Kraemer}, \citenamefont {Sailer}, \citenamefont {Tu},
  \citenamefont {Weigel}, \citenamefont {Wolf}, \citenamefont {{Crespo
  L\'{o}pez-Urrutia}},\ and\ \citenamefont {Blaum}}]{Sturm2019}%
  \BibitemOpen
  \bibfield  {author} {\bibinfo {author} {\bibfnamefont {S.}~\bibnamefont
  {Sturm}}, \bibinfo {author} {\bibfnamefont {I.}~\bibnamefont {Arapoglou}},
  \bibinfo {author} {\bibfnamefont {A.}~\bibnamefont {Egl}}, \bibinfo {author}
  {\bibfnamefont {M.}~\bibnamefont {H\"{o}cker}}, \bibinfo {author}
  {\bibfnamefont {S.}~\bibnamefont {Kraemer}}, \bibinfo {author} {\bibfnamefont
  {T.}~\bibnamefont {Sailer}}, \bibinfo {author} {\bibfnamefont
  {B.}~\bibnamefont {Tu}}, \bibinfo {author} {\bibfnamefont {A.}~\bibnamefont
  {Weigel}}, \bibinfo {author} {\bibfnamefont {R.}~\bibnamefont {Wolf}},
  \bibinfo {author} {\bibfnamefont {J.}~\bibnamefont {{Crespo
  L\'{o}pez-Urrutia}}}, \ and\ \bibinfo {author} {\bibfnamefont
  {K.}~\bibnamefont {Blaum}},\ }\href {\doibase 10.1140/epjst/e2018-800225-2}
  {\bibfield  {journal} {\bibinfo  {journal} {Eur. Phys. J. Spec. Top.}\
  }\textbf {\bibinfo {volume} {227}},\ \bibinfo {pages} {1425} (\bibinfo {year}
  {2019})}\BibitemShut {NoStop}%
\bibitem [{\citenamefont {{Soria Orts}}\ \emph {et~al.}(2007)\citenamefont
  {{Soria Orts}}, \citenamefont {{Crespo L{\'o}pez-Urrutia}}, \citenamefont
  {Bruhns}, \citenamefont {{Gonz{\'a}lez Mart{\'i}nez}}, \citenamefont
  {Harman}, \citenamefont {Jentschura}, \citenamefont {Keitel}, \citenamefont
  {Lapierre}, \citenamefont {Tawara}, \citenamefont {Tupitsyn} \emph
  {et~al.}}]{Soriaorts2007}%
  \BibitemOpen
  \bibfield  {author} {\bibinfo {author} {\bibfnamefont {R.}~\bibnamefont
  {{Soria Orts}}}, \bibinfo {author} {\bibfnamefont {J.~R.}\ \bibnamefont
  {{Crespo L{\'o}pez-Urrutia}}}, \bibinfo {author} {\bibfnamefont
  {H.}~\bibnamefont {Bruhns}}, \bibinfo {author} {\bibfnamefont {A.~J.}\
  \bibnamefont {{Gonz{\'a}lez Mart{\'i}nez}}}, \bibinfo {author} {\bibfnamefont
  {Z.}~\bibnamefont {Harman}}, \bibinfo {author} {\bibfnamefont {U.~D.}\
  \bibnamefont {Jentschura}}, \bibinfo {author} {\bibfnamefont {C.~H.}\
  \bibnamefont {Keitel}}, \bibinfo {author} {\bibfnamefont {A.}~\bibnamefont
  {Lapierre}}, \bibinfo {author} {\bibfnamefont {H.}~\bibnamefont {Tawara}},
  \bibinfo {author} {\bibfnamefont {I.~I.}\ \bibnamefont {Tupitsyn}},  \emph
  {et~al.},\ }\href@noop {} {\bibfield  {journal} {\bibinfo  {journal} {Phys.
  Rev. A}\ }\textbf {\bibinfo {volume} {76}},\ \bibinfo {pages} {052501}
  (\bibinfo {year} {2007})}\BibitemShut {NoStop}%
\end{thebibliography}
\end{document}